\documentclass[11pt,twoside,a4paper]{article}
\pdfoutput=1

\usepackage[framemethod=tikz]{mdframed}
\usepackage{amsmath, amsfonts, amssymb, amsthm}
\usepackage{graphicx}
\usepackage{float}
\usepackage{dsfont}
\usepackage{color}
\usepackage{cite}

\usepackage[width=\textwidth,font=small,labelfont=bf,
labelsep=endash]{caption}
\usepackage{titlesec}

\titleformat{\section}
  {\normalfont\fontsize{13}{16}\bfseries}{\thesection}{1em}{}
\titleformat{\subsection}
  {\normalfont\fontsize{11}{14}\bfseries}{\thesubsection}{1em}{}

\normalsize
\newcommand\Tr {\mathrm{Tr}}

\newcommand\re {\mathrm{Re}}

\numberwithin{equation}{section}
\makeatletter
\newcommand{\vast}{\bBigg@{4}}
\newcommand{\Vast}{\bBigg@{5}}
\makeatother

\begin{document}

\title{Entanglement in $(1+1)$-dimensional Free Scalar Field Theory: Tiptoeing between Continuum and Discrete Formulations}
\author{Dimitrios Katsinis$^{1,2}$, Georgios Pastras$^{2,3}$}
\date{$^1$Instituto de F\'isica, Universidade de S\~ao Paulo, Rua do Mat\~ao Travessa 1371, 05508-090 S\~ao Paulo, SP, Brazil\\
$^2$National and Kapodistrian University of Athens,\\ Department of Physics, 15784 Zografou, Attiki, Greece\\
$^3$Laboratory for Manufacturing Systems and Automation, Department of Mechanical Engineering and Aeronautics,\\ University of Patras, 26504 Rio, Patra, Greece
\linebreak\\
\texttt{dkatsinis@phys.uoa.gr, pastras@lms.mech.upatras.gr}}

\vskip .5cm

\maketitle

\begin{abstract}
We review some classic works on ground state entanglement entropy in $(1+1)$-dimensional free scalar field theory. We point out identifications between the methods for the calculation of entanglement entropy and we show how the formalism developed for the discretized theory can be utilized in order to obtain results in the continuous theory. We specify the entanglement spectrum and we calculate the entanglement entropy for the theory defined on an interval of finite length $L$. Finally, we derive the modular Hamiltonian directly, without using the modular flow, via the continuum limit of the expressions obtained in the discretized theory. In a specific coordinate system, the modular Hamiltonian assumes the form of a free field Hamiltonian on the Rindler wedge.
\end{abstract}

\newpage

\tableofcontents

\newpage

\section{Introduction\protect\label{sec:intro}}

Entanglement \cite{Einstein:1935rr,schrodinger_1935} is a property of quantum systems without a classical analogue. Let us consider two quantum systems that are part of a larger composite system. When these subsystems are entangled, the results of local measurements performed on them are correlated. The converse is not generally true, since the measurements may be affected by classical correlations. Whenever the overall system lies in a pure state there is a simple measure of entanglement, the so called entanglement entropy\footnote{It is natural to wonder how classical correlations and correlations due to quantum entanglement can be distinguished for mixed states. The answer to this question is complicated and such an analysis is beyond our scope. The reader may consult \cite{Chruscinski:2014oca}.}. The entanglement entropy is defined as the von Neumann entropy of the reduced density matrix. The latter, which describes subsystem $A$, is obtained from the density matrix via tracing out the degrees of freedom of the complementary subsystem $A^C$, i.e.
\begin{equation}
\rho_A=\Tr_{A^C}\left[\rho\right].
\end{equation}
Then, the entanglement entropy of the system $A$ is defined as
\begin{equation}
S_A:=-\Tr_{A}\left[\rho_A\ln\rho_A\right].
\end{equation}
Entanglement entropy has many applications. For example, it is related to quantum phase transitions \cite{Vidal:2002rm}, as well as to the topological order \cite{Levin:2006zz,Kitaev:2005dm}.

%%%%%% Methodologies

% Discretization
The study of entanglement in free Quantum Field Theory was initiated a long time ago \cite{Sorkin:1984kjy,Bombelli:1986rw,Srednicki:1993im}. These works rely on the study in position representation of a quantum mechanical system, which is obtained after the lattice discretization of the Quantum Field Theory. The basic finding of these works is the following: for a free scalar field at its ground state, considering the degrees of freedom of some spatial region as a subsystem, the entanglement entropy is proportional to the area of this region and not to the volume, as one would naively expect. Remarkably, this behaviour resembles the area law obeyed by the entropy of Black Holes \cite{Bekenstein:1973ur,Bardeen:1973gs}.

% Replica Trick
Another method for the calculation of entanglement entropy was introduced in \cite{Callan:1994py,Holzhey:1994we}, where a Euclidean-time/path-integral approach is followed.\footnote{See also \cite{Larsen:1994yt} for a discussion on fermions.} The traces of powers of the reduced density matrix $\Tr[\rho_A^n]$, which are essentially the so-called R\'enyi entanglement entropies \cite{renyi1961}, are calculated, and the entanglement entropy is obtained by analytic continuation, as the limit
\begin{equation}
S_{A}=\lim_{n\rightarrow 1^+}\frac{\Tr[\rho_A^n]-1}{1-n}=-\frac{d}{dn}\Tr[\rho_A^n]\Big\vert_{n=1}.
\end{equation}
This method came to be known as the ``replica trick" and was further developed for CFTs in \cite{Calabrese:2004eu}. The study of entanglement in Quantum Field Theory became a very active field on its own, probing many interesting properties of field theories such as the renormalization group flow \cite{Casini:2004bw,Balasubramanian:2011wt}. The reader can consult the reviews \cite{Calabrese:2009qy,Casini:2009sr,Casini:2022rlv}, along with \cite{Amico:2007ag,Eisert:2008ur}, which focus on discrete systems.

% Entanglement in AdS/CFT
Entanglement entropy is also an important line of research in the framework of the AdS/CFT correspondence \cite{Maldacena:1997re,Gubser:1998bc,Witten:1998qj}. Ruy and Takayanagi proposed a prescription for the calculation of holographic entanglement entropy \cite{Ryu:2006bv,Ryu:2006ef}. This prescription enables the study of entanglement in the non-perturbative regime of theories with holographic duals, which would be impossible directly. The prescription was generalized for time-dependent setups \cite{Hubeny:2007xt} and was proven in \cite{Lewkowycz:2013nqa}.
%The first quantum correction to this formula is calculated in \cite{Faulkner:2013ana}, whereas a quantum-mechanically exact prescription is suggested in \cite{Engelhardt:2014gca}. 
An impressive outcome of this program is the derivation of the Page curve of an evaporating Black Hole \cite{Page:1993wv} in the framework of AdS/CFT \cite{Penington:2019npb,Almheiri:2019psf}. Another compelling finding concerns the use of holographic entanglement entropy as an order parameter for confinement \cite{Klebanov:2007ws}. The reader may consult \cite{Nishioka:2009un,Takayanagi:2012kg,VanRaamsdonk:2016exw,Rangamani:2016dms,Nishioka:2018khk} for an overview of this vast subject.

% Comparison
The above three methods, namely the one relying on discretization, which in the following will be referred to as ``Srednicki's method", the replica trick and the holographic calculations are the most common techniques for the calculation of entanglement entropy. The first one requires numerical calculations, thus obtaining results in an analytic form is compelling. On the other hand, it provides straightforwardly not only the entanglement entropy but also the spectrum of the reduced density matrix, which is known as the entanglement spectrum, and its eigenfunctions. We are particularly interested in the calculation of the latter in order to probe the relation between entanglement and gravity. 

% Emergent Gravity
A statistical/emergent interpretation of gravity was considered long ago \cite{Jacobson:1995ab}, see also \cite{Verlinde:2010hp}. It was shown that whenever the entropy of horizons is proportional to their area and the entropy obeys the first law of thermodynamics, then the metric satisfies the Einstein's equations of general relativity. Subsequent works point out that there is a quantum version of these arguments \cite{Jacobson:2012yt,Jacobson:2015hqa}.

Van Raamsdonk put forth the idea that AdS/CFT implies that entanglement and gravitation are intertwined \cite{VanRaamsdonk:2010pw,VanRaamsdonk:2009ar}. This idea became more concrete in subsequent works \cite{Lashkari:2013koa,Faulkner:2013ica}. In the same spirit, Maldacena and Susskind proposed that entanglement is the analogue of gravitational wormholes, a statement that goes by the name ER=EPR \cite{Maldacena:2013xja}. Susskind took a step further and argued in an open letter that we already know that General Relativity and Quantum Mechanics are exactly the same thing \cite{Susskind:2017ney}. In his words: ``\emph{Where there is quantum mechanics, there is also gravity}". 

A concrete realization of such a mechanism for emergent gravity has been developed in the context of AdS/CFT. Any quantum subsystem obeys the first law of entanglement thermodynamics, namely 
\begin{equation}
\delta S_{A}=\delta \left\langle H_A\right\rangle,
\end{equation}
where $H_A$ is the modular Hamiltonian that is defined as $\rho_{A}=e^{-H_A}$. It was shown that the first law of entanglement thermodynamics in the bulk is equivalent to the linearized Einstein equations over the AdS backround \cite{Lashkari:2013koa,Faulkner:2013ica}, see also \cite{Blanco:2013joa}.

% Emergent Gravity from Entanglement
In the spirit of the above, it is worth probing these considerations beyond the framework AdS/CFT. One way to approach these considerations is the following: The entanglement entropy of a scalar field obeys an area law at the vacuum state,\footnote{Keep in mind that the vacuum is a very special state. For an arbitrary state entanglement entropy is extensive \cite{Page:1993df}.} which is a striking similarity with the entropy of Black Holes. Is there any other signature of gravitational physics in such a system? After all entanglement entropy is only a particular measure of entanglement. The spectrum and the eigenfunctions of the reduced density matrix contain much more information.\footnote{We already know that this is the case for instance when it comes to the Quantum Hall Effect \cite{Moore:1991ks,Li:2008kda}.}

Given the spectral decomposition of the reduced density matrix one can determine the modular Hamiltonian, which seems to play an important role in the relation between entanglement and gravity, at least in the context of AdS/CFT. For continuous systems the modular Hamiltonian is calculated using the modular flow \cite{Bisognano:1975ih,Bisognano:1976za,Casini:2011kv,Hislop:1981uh}, but this tool is not always applicable. For this reason we demonstrate how the modular Hamiltonian can be derived via the continuum limit of Srednicki's method.

% Modular Hamiltonian - Correlation Functions 
Even for a free scalar field it is not always possible to calculate the entanglement entropy. For harmonic systems, there is a special class of states of the composite system which at any time instant have the property that both the density matrix and the reduced density matrix are Gaussian in position representation, making the calculations possible. In general, \emph{at a given time instant}, the entanglement entropy depends only on the \emph{state} and not on the \emph{dynamics} of the system. Obviously the latter determines the time evolution of the state and thus of the entanglement entropy. These Gaussian states are: 
\begin{enumerate}
\item the ground state \cite{Bombelli:1986rw,Srednicki:1993im,Callan:1994py,Holzhey:1994we,Katsinis:2017qzh}
\item thermal states \cite{Katsinis:2019vhk,Katsinis:2019lis}
\item coherent states, which share the same entanglement spectrum with the vacuum \cite{Benedict:1995yp,Katsinis:2022fxu}. As a result, the time evolution of the reduced system is unitary.
\item squeezed states \cite{Bianchi:2015fra,Katsinis:2023hqn,Katsinis:2024sek}, whose entanglement is time dependent.
\end{enumerate}
These states have another interesting property. The modular Hamiltonian describes a non-interacting theory, i.e. one can define normal modes on the subsystem. 

For free scalar field theory at its vacuum state, when tracing out half of the space, one can calculate the modular Hamiltonian using the mapping from the Minkowski to the Rindler space. The time evolution is generated by the boost generator, which coincides with the modular Hamiltonian \cite{Bisognano:1976za,Bisognano:1975ih}. In general, the latter is a non-local function of the stress-energy tensor, but for this setup it is local. Due to this fact, the modular Hamiltonian can be calculated via the modular flow \cite{Casini:2011kv,Hislop:1981uh}. The locality of the modular flow manifests itself in the fact that the modular Hamiltonian is a weighted integral of the energy density. This subject is further discussed in \cite{Cardy:2016fqc,Arias:2016nip,Arias:2017dda,Longo:2020amm}.

Gaussian density matrices have another interesting property. The correlation functions split to products of two-point functions. Using this fact one can relate the eigenvalues of the modular Hamiltonian to the two-point correlation functions \cite{Peschel_2003}, see also the appendix E of \cite{Katsinis:2023hqn} for the application of this method in the case of squeezed states. An alternative approach is based on a generalized/covariant definition of entanglement entropy that is naturally related to Wightman functions and is applicable to Gaussian states \cite{Sorkin:2012sn}. Applications for continuum and discrete settings are presented in \cite{Saravani:2013nwa,Sorkin:2016pbz}. 

In this work we revisit the calculation of entanglement entropy in (1+1)~-dimensional free scalar field theory. Initially, we review the distinct approaches for its calculation, namely Srednicki's method, the replica trick and the correlation functions method. We point out the relations between them. More specifically we show that Srednicki's method and the correlation functions method are equivalent. Then we show that for a massless scalar field, the continuum limit of these methods gives rise to the results obtained using the replica trick \cite{Callan:1994py}. By exploiting the correspondence between the aforementioned methods we derive the entanglement spectrum in the continuous theory for a system of finite length. In the infinite-size limit and when the subsystem is the halfspace, the spectrum was derived in \cite{Callan:1994py}, even though this fact was not explicitly stated. Finally, using this twofold approach, we tackle a long-standing problem, which has recently been discussed \cite{DiGiulio:2019cxv,Eisler:2020lyn} using numerical methods, namely the derivation of the modular Hamiltonian of the continuous (1+1)-dimensional massless scalar field theory starting from the discretized theory, see also \cite{Arias:2018tmw}. The reader interested in the implications of the boundary conditions may consult \cite{Estienne:2023ekf}. A key element in our derivation is a coordinate transformation that enables the systematic calculation of the entanglement spectrum. This transformation is directly related to that used in the implementation of the replica trick in the CFT calculation.

The structure of the paper is as follows: In Section \ref{sec:Srednicki_review} we review the calculation of entanglement entropy for a system of harmonic oscillators. In Section \ref{sec:Continuous_Theory} we briefly review the discretized (1+1)-dimensional scalar field theory. We present the entanglement spectrum and the matrices required for the calculations. Then, we derive the entanglement spectrum of the continuous theory, along with the corresponding continuous kernels. In Section \ref{sec:Finite_Interval} we derive the eigenvalues and the eigenfunctions of the reduced density matrix, the entanglement entropy and the modular Hamiltonian. Finally, in Section \ref{sec:discussion} we discuss our results. There is a series of appendices containing details of the calculations.

\section{Entanglement of a Bosonic Harmonic System\protect\label{sec:Srednicki_review}}

In this section we review Srednicki's approach for the calculation of entanglement entropy in a harmonic bosonic system at its vacuum state \cite{Srednicki:1993im}. We show that this method is equivalent to the correlation functions method introduced by Peschl \cite{Peschel_2003}. 
\subsection{Srednicki's Method}
We consider a bosonic harmonic  system, with  dynamics determined by the Hamiltonian
\begin{equation}
H=\frac{1}{2}P_X^T P_X+\frac{1}{2}X^T K X,
\end{equation}
where $K$ is a positive-definite symmetric matrix. The vector $X$ contains the positions of the degrees of freedom, whereas the vector $P_X$ contains the conjugate momenta. One can introduce normal coordinates $Y=OX$, where $O$ is an orthogonal matrix, so that $K=O^T K_D O$ with $K_D$ a diagonal matrix. In terms of these coordinates the Hamiltonian assumes the form 
\begin{equation}
H=\frac{1}{2}P^T_Y P_Y+\frac{1}{2}Y^T K_D Y.
\end{equation}
We also introduce the positive-definite matrix $\Omega_D$ that obeys $K_D=\Omega_D^2$. The elements of $\Omega_D$ are the eigenfrequencies of the system. The ground state wave function is given by
\begin{equation}
\Psi=\left(\frac{\det\Omega_D}{\pi^N}\right)^{1/4}\exp\left(-\frac{1}{2}Y^T\Omega_D Y\right) =\left(\frac{\det\Omega}{\pi^N}\right)^{1/4}\exp\left(-\frac{1}{2}X^T\Omega X\right).
\end{equation}
In position representation the density matrix, which is defined as $\rho(X,X^\prime)=\Psi(X)\Psi^*(X^\prime)$, reads
\begin{equation}
\rho(X,X^\prime)\hspace{-0.03cm}=\hspace{-0.03cm}\frac{\sqrt{\det\Omega}}{\pi^{\frac{N}{2}}}\exp\left[-\frac{1}{2}\left(X^T\Omega X+X^{\prime T}\Omega X^\prime\right)\right].
\end{equation}
We trace out the oscillators $1$ to $n$. We consider the following blocks of the matrix $\Omega$, along with the coordinate vectors $X$ and $X^\prime$:
\begin{equation}
\Omega=\begin{pmatrix}
\Omega_A & \Omega_B\\
\Omega_B^T & \Omega_C
\end{pmatrix},\quad X=\begin{pmatrix}
x_c\\
x
\end{pmatrix},\quad X^\prime=\begin{pmatrix}
x^\prime_c\\
x^\prime
\end{pmatrix}.
\end{equation}
Naturally, the block $\Omega_A$ is an $n\times n$ matrix, $\Omega_C$ is an $(N-n)\times(N- n)$ one and so on.

It is trivial to calculate the reduced density matrix by setting $x^\prime_c=x_c$ and integrating over $x_c$. Doing so, we obtain
\begin{equation}\label{eq:rho_red_g_b}
\rho_{\textrm{red}}(x,x^\prime)=\sqrt{\frac{\det\left(\gamma-\beta\right)}{\pi^{N-n}}}\exp\left[-\frac{1}{2}\left(x^T\gamma x+x^{\prime T}\gamma x^\prime\right)+x^T\beta x\right],
\end{equation}
where the matrices $\gamma$ and $\beta$ read
\begin{equation}\label{eq:gamma_beta_def}
\gamma=\Omega_C-\frac{1}{2}\Omega_B^T\Omega_A^{-1}\Omega_B,\quad \beta=\frac{1}{2}\Omega_B^T\Omega_A^{-1}\Omega_B.
\end{equation}
Notice that both $\gamma$ and $\beta$ are symmetric matrices. Using appropriate changes of variables, it can be shown \cite{Srednicki:1993im} that the spectrum f the reduced density matrix has the form
\begin{equation}\label{eq:spec}
p_{\vec{n}}=\prod_{i=1}^{N-n}\left(1-\xi_i\right)\xi_i^{n_i},
\end{equation}
where $\xi_i$ are the eigenvalues of the matrix $\Xi$, which is defined in terms of the matrix $\gamma^{-1}\beta$ as
\begin{equation}\label{eq:definition_Xi}
\Xi=\frac{\gamma^{-1}\beta}{I+\sqrt{I-\left(\gamma^{-1}\beta\right)^2}}.
\end{equation}
The spectrum \eqref{eq:spec} is quasi-thermal. The modes of the subsystem lie in thermal states, but these states do not correspond to a common temperature. It follows that the eigenvalues $\xi_i$ of the matrix $\Xi$ are determined by the eigenvalues of the matrix $\gamma^{-1}\beta$. The entanglement entropy $S_{\textrm{EE}}$ is given in term of $\xi_i$ by
\begin{equation}\label{eq:SEE_sum}
S_{\textrm{EE}}=\sum_{i=1}^{N-n}\left[-\ln(1-\xi_i)-\frac{\xi_i}{1-\xi_i}\ln\xi_i\right].
\end{equation}

It is easy to show that the matrix $\Xi$ can also be expressed in the form
\begin{equation}\label{eq:xi_sec}
\Xi=\frac{\sqrt{\frac{I+\gamma^{-1}\beta}{I-\gamma^{-1}\beta}}-I}{\sqrt{\frac{I+\gamma^{-1}\beta}{I-\gamma^{-1}\beta}}+I},
\end{equation}
which is crucial in order to establish the connection between Srednicki's method and the correlation matrix method.
\subsection{Entanglement Entropy \& Correlation Functions}
In this section we relate the formulae of the previous section to the correlation functions of the positions and the momenta of the system. The sum and the difference of the matrices $\gamma$ and $\beta$, defined in \eqref{eq:gamma_beta_def}, have very simple expressions, namely
\begin{equation}
\gamma+\beta=\Omega_C,\quad \gamma-\beta=\Omega_C-\Omega_B^T\Omega_A^{-1}\Omega_B.
\end{equation}
Notice that $\gamma-\beta$ is actually the Schur complement of $\Omega_C$. In other words, denoting the blocks of the matrix $\Omega^{-1}$ as
\begin{equation}
\begingroup
\renewcommand*{\arraystretch}{1.5}
\Omega^{-1}=\begin{pmatrix}
\left(\Omega^{-1}\right)_A & \left(\Omega^{-1}\right)_B\\
\left(\Omega^{-1}\right)^T_B & \left(\Omega^{-1}\right)_C
\end{pmatrix}
\endgroup
\end{equation}
it follows that
\begin{equation}
\gamma-\beta=\left(\left(\Omega^{-1}\right)_C\right)^{-1}.
\end{equation}

Defining the matrix $M$ via the equation 
\begin{equation}\label{eq:mat_M_definition}
M:=\left(\Omega^{-1}\right)_C\Omega_C,
\end{equation}
the matrix $\gamma^{-1}\beta$ can be expressed in terms of the matrix $M$ as
\begin{equation}
\gamma^{-1}\beta=\frac{M-I}{M+I}.
\end{equation}
Interestingly enough, using \eqref{eq:xi_sec} the matrix $\Xi$ also assumes a very simple form, namely
\begin{equation}\label{eq:Xi_of_M}
\Xi=\frac{\sqrt{M}-I}{\sqrt{M}+I}.
\end{equation}
It follows that the spectrum of the reduced density matrix is determined solely by the matrix $M$. This is the famous prescription for the calculation of entanglement entropy in terms of correlation functions \cite{Peschel_2003}, see also \cite{Sorkin:2012sn}. Indeed, it is straightforward to show that
\begin{align}
{\mathcal{X}}_{ij}:=\left\langle X_i X_j\right\rangle&=\frac{1}{2}\left(\Omega^{-1}\right)_{ij},\label{eq:correlations_X}\\ 
{\mathcal{P}}_{ij}:=\left\langle P_i P_j\right\rangle&=\frac{1}{2} \Omega_{ij},\label{eq:correlations_P}\\
 \left\langle X_i P_j\right\rangle=-\left\langle P_j X_i\right\rangle&=\frac{i}{2}\delta_{ij}.
\end{align}
As a result, we conclude that $M=4{\mathcal{X}}{\mathcal{P}}$.

Equation \eqref{eq:mat_M_definition} implies that the entanglement spectrum is determined solely by the blocks of the matrices $\Omega$ and $\Omega^{-1}$, which correspond to the subsystem under consideration. Even though these matrices refer to the overall system, their $C-$blocks determine the correlation functions of the degrees of freedom and the conjugate momenta in the subsystem under consideration. As a consequence, equation \eqref{eq:Xi_of_M} connects Srednicki's method to the correlation functions method.

The idea behind the calculation of entanglement entropy using correlation functions is the following: Since the ground state is Gaussian, $n$-point correlation functions factorize into $2$-point correlation functions, in accordance with Wick's theorem. Postulating that the modular Hamiltonian describes a harmonic theory and further demanding that it reproduces the correct 2-point functions, determines it. More specifically, one assumes that the reduced density matrix is written in the form
\begin{equation}
\rho_\textrm{red}=\mathcal{N} \exp\left(-H_M\right),\label{eq:Modular_Modes_rho}
\end{equation}
where $H_M$ is the modular Hamiltonian, given by
\begin{equation}
H_M=\sum_{i=1}^{N-n}\epsilon_i a^\dagger_i a_i\label{eq:Modular_Modes_HM}\\
\end{equation}
and $\mathcal{N}$ is the normalization factor
\begin{equation}
\mathcal{N} =\prod_{i=1}^{N-n}\left(1-e^{-\epsilon_i}\right)\label{eq:Modular_Modes_N}
\end{equation}
that ensures $\Tr \rho_\textrm{red}=1$. One can show that the eigenvalues of the matrix $M$, that we denoted as $M_i$, are related to the parameters $\epsilon_i$ as
\begin{equation}\label{eq:Modular_Energy}
\epsilon_i=2\coth^{-1}\left(\sqrt{M_i}\right)=-\ln\xi_i,
\end{equation}
where $\xi_i$ are the eigenvalues of matrix $\Xi$. The last equality follows directly from equation \eqref{eq:Xi_of_M}. See \cite{Casini:2009sr} for more details.
\subsection{Eigenstates of the Reduced Density Matrix\protect\label{subsec:eigenstates_main}}
So far we have focused on the entanglement spectrum. In his seminal paper Srednicki derived it via a series of coordinate transformations, namely, a rotation that diagonalizes $\gamma$, a rescaling that sets $\gamma$ to identity and one more rotation that diagonalizes the matrix 
\begin{equation}\label{eq:beta_hat}
\hat{\beta}=\gamma^{-1/2}\beta \gamma^{-1/2}.
\end{equation}
In an obvious manner, the matrix $\hat{\beta}$ has the same eigenvalues with the matrix $\gamma^{-1}\beta$. The eigenfunctions of the reduced density matrix read
\begin{equation}
\Psi_{\vec{n}}(\vec{\hat{x}})=\prod_{i=1}^{N-n} \phi_{n_i}(\hat{x}_i),
\end{equation}
where $\hat{x}_i$ are the coordinates that diagonalize the matrix $\hat{\beta}$ and $\phi_{n}(x)$ is the wave function of the $n-$th excited state of the harmonic oscillator with eigenfrequency $\alpha_i=\sqrt{1-\hat{\beta}^2_i}$, where $\hat{\beta}_i$ are the eigenvalues of $\hat{\beta}$, namely\footnote{There is a simple way to expand the reduced density matrix in terms of its eigenfunctions. After the coordinate transformations the reduced density matrix has the form
\begin{equation}
\rho_{\textrm{red}}\left(\hat{x},\hat{x}^\prime\right)=\prod_{i=n}^N \sqrt{\frac{1-\hat{\beta}_i}{\pi}}\exp\left[-\frac{1}{2}\left(\hat{x}_i^2+\hat{x}_i^{\prime 2}\right)+\hat{x}_i^T\hat{\beta}\hat{x}_i^\prime\right].
\end{equation}
Then, using the Melher's formula
\begin{equation}
\frac{1}{\sqrt{1-\sigma^2}}\exp\left[-\frac{\sigma^2\left(x^2+y^2\right)-2\sigma xy}{1-\sigma^2}\right]=\sum_{n=0}^\infty \frac{\left(\sigma/2\right)^n}{n!}H_n(x)H_n(y),
\end{equation}
with $x=\sqrt{\alpha_i}\hat{x}_i$, $y=\sqrt{\alpha_i}\hat{x}_i^\prime$ and $\sigma=\xi_i$, where $\xi_i$ is defined in \eqref{eq:xi_i_def}, we obtain the expansion of the reduced density matrix to its eigenfunctions.}
\begin{equation}\label{eq:nth_wave_function}
\phi_{n_i}(\hat{x}_i)=\sqrt[4]{\frac{\alpha_i}{\pi}}\frac{H_{{n_i}}\left(\sqrt{\alpha_i}\hat{x}_i\right)}{\sqrt{n_i!2^{n_i}}}\exp\left(-\frac{\alpha_i}{2}\hat{x}_i^2\right).
\end{equation}
The above coordinate transformations include a non-trivial rescaling, which messes up the normalization of the reduced density matrix. As a consistency check in Appendix \ref{sec:Algebraic} we derive the spectrum and the eigenfunctions using creation and annihilation operators.
\subsection{The Modular Hamiltonian\protect\label{sec:Mudular_2d}}
The calculation of the previous section implies that the reduced density matrix can be expressed in terms of the creation and annihilation operators associated to the wave functions of equation \eqref{eq:nth_wave_function} as
\begin{equation}
\hat{\rho}_{\textrm{red}}= \det(I-\Xi)\exp\left(-\hat{H}_M\right),
\end{equation}
where the modular Hamiltonian $\hat{H}_M$ is given by
\begin{equation}
\hat{H}_M=-\sum_{i=1}^{N-n}\ln(\xi_i)\hat{A}^\dagger_i \hat{A}_i.
\end{equation}
The explicit forms of the operators $\hat{A}_i$ and $\hat{A}^\dagger_i$ are given by \eqref{eq:op_A} and \eqref{eq:op_A_dag}, respectively. The modular Hamiltonian $\hat{H}_M$ describes a free theory. It is a matter of algebra to show that it may be written  as a differential operator in the form
\begin{equation}\label{eq:modular_disc}
\hat{H}_M=\sum_{m,\ell}\left(-\tilde{K}_{m\ell}\frac{\partial}{\partial x_m}\frac{\partial}{\partial x_\ell}+\tilde{V}_{m\ell}x_m x_\ell\right) +\frac{1}{2}\ln(\det\Xi),
\end{equation}
where the matrices $\tilde{K}$ and $\tilde{V}$ are defined as
\begin{align}
\tilde{K}_{m \ell}&=-\sum_{i=1}^{N-n}\frac{v^{i}_{\rho}v^{i}_{k}}{\alpha_i}\left(\gamma^{-1}\right)_{\rho \ell}\left(\gamma^{-1}\right)_{k m}\ln\xi_i,\\ \tilde{V}_{m\ell}&=-\frac{1}{2}\sum_{i=1}^{N-n}\alpha_iv^{i}_{m}v^{i}_{\ell}\ln\xi_i.
\end{align}
Using \eqref{eq:eig_normalization} and \eqref{eq:beta_decomposition}, it turns out that the matrices $\tilde{K}$ and $\tilde{V}$ read
\begin{align}
\tilde{K}&=-\frac{1}{2}\frac{\ln\Xi}{\sqrt{I-\left(\gamma^{-1}\beta\right)^2}}\gamma^{-1},\\
\tilde{V}&=-\frac{1}{2}\gamma\sqrt{I-\left(\gamma^{-1}\beta\right)^2}\ln\Xi.
\end{align}
The above expressions assume a much simpler form in terms of the matrix $M$, defined in \eqref{eq:mat_M_definition}, namely
\begin{align}
\tilde{K}&=\frac{\mathrm{arccoth}\sqrt{M}}{\sqrt{M}}\left(\Omega^{-1}\right)_C,\label{eq:modular_K}\\
\tilde{V}&=\Omega_C\frac{\mathrm{arccoth}\sqrt{M}}{\sqrt{M}}.\label{eq:modular_V}
\end{align}
Bearing in mind that the matrix $M$ is expressed in terms of $\Omega_C$ and $\left(\Omega^{-1}\right)_C$, the modular Hamiltonian is determined by the 2-point correlation functions of the subsystem under consideration, as expected.
\section{(1+1)-dimensional Field Theory\protect\label{sec:Continuous_Theory}}
\subsection{The Discrete Theory}
In the previous section we have shown that the calculation of the spectrum of the reduced density matrix is equivalent to the calculation of the eigenvalues and eigenvectors of the matrix $M=\left(\Omega^{-1}\right)_C\Omega_C$. So, given a Hamiltonian $H=\frac{1}{2}P^T P+\frac{1}{2}X^T K X$, where the coupling matrix $K$ is positive definite, we are interested in finding the matrix $\Omega$ that is positive-definite and obeys that $K=\Omega^2$. 

We focus on (1+1)-dimensional free real scalar field theory defined on an interval of length $L$. In order to apply the formalism that we developed for a system of coupled harmonic oscillators, we have to discretize the degrees of freedom. Introducing a homogeneous lattice, we obtain 
\begin{equation}
H=\frac{1}{2}\sum_{j=1}^N\left[\pi_j^2+\frac{1}{a^2}\left(\varphi_{j+1}-\phi_j\right)^2+\mu^2\varphi_j^2\right],
\end{equation}
where $a$ is the lattice spacing. We used the scheme
\begin{equation}\label{eq:disc_scheme}
\int_0^\infty dx\rightarrow a\sum_{j=1}^{N},\quad x=ja,\quad L=(N+1)a,\quad \varphi(x)\rightarrow\frac{\phi_j}{\sqrt{a}},\quad \pi(x)\rightarrow\frac{\pi_j}{\sqrt{a}}.\phantom{iqnfeiqnf}
\end{equation}
This Hamiltonian describes a system of coupled harmonic oscillators. The coupling matrix $K$ of the system has a very simple form, namely
\begin{equation}\label{eq:matK_2d}
K_{ij}=\frac{1}{a^2}\left[\left(2+\mu^2a^2\right)\delta_{i,j}-(\delta_{i,j+1}+\delta_{i+1,j})\right].
\end{equation}
This is a Toplitz tridiagonal matrix with known eigenvalues and eigenvectors. Its spectral decomposition is
\begin{equation}
K_{ij}=\frac{2}{N+1}\sum_{k=1}^{N}w_k^2\sin\frac{i k \pi}{N+1}\sin\frac{j k \pi}{N+1},
\end{equation}
where the eigenvalues $w_k^2$ are given by
\begin{equation}
w_k^2=\mu^2+\frac{4}{a^2}\sin^2\frac{\pi k}{2(N+1)}.
\end{equation}
\subsection{Back to the Continuum}
So far we have taken advantage of the discretization in order to deal with a well-posed problem, namely the calculation of the entanglement entropy of a system of quantum harmonic oscillators. The formalism developed in Section \ref{sec:Srednicki_review} is directly applicable to the discretized theory. Unfortunately, we are not able to proceed much further in this framework. We will work in the continuum limit of the expressions that were obtained in the discretized theory. This way the discrete theory resolves the ambiguities of the definition of entanglement entropy that emerge when we study directly the continuous theory. 

In the continuum limit the discrete indices become continuous variables, the column vectors become functions of these continuous variables and the matrices become bilinear kernels. Essentially we have to invert the discretization procedure appearing in equation \eqref{eq:disc_scheme}. Regarding the size of the overall system, the continuum limit corresponds to $a\rightarrow0$, $N\rightarrow\infty$, so that $L=(N+1)a$ is fixed. Similarly, for the rest of the variables it corresponds to $a\rightarrow0$, $i\rightarrow\infty$, so that $x=i a$ is fixed and so on. Regarding the bilinear kernels the continuum limit is
\begin{equation}\label{eq:kernel_limit}
K(x,x^\prime)=\lim_{a\rightarrow0}\frac{1}{a}K_{ij},
\end{equation}
where $x$ and $x^\prime$ are the continuous coordinates associated to $i$ and $j$ respectively. The factor of $1/a$ is inserted because the sums over the indices become integrals over a continuous variable. This factor is required in order to have the correct measure of integration. 

It turns out that
\begin{equation}\label{eq:kernel_K}
K(x,x^\prime)=\frac{2}{L}\sum_{k=1}^{\infty}\omega_k^2\sin\frac{k \pi x}{L}\,\sin\frac{k \pi x^\prime}{L},
\end{equation}
where the eigenvalues $\omega_k^2$ are given by
\begin{equation}
\omega_k^2=\mu^2+\frac{\pi^2}{L^2}k^2,
\end{equation}
while $x$ and $x^\prime$ are valued in $\left[0,L\right]$ \footnote{One may wonder about the behaviour of the sum over $k$. The upper bound of summation is $N$ and we use the limit
\begin{equation}
\lim_{a\rightarrow 0}\frac{4}{a^2}\sin^2\frac{a \pi k}{2L}=\left(\frac{\pi k}{L}\right)^2,
\end{equation}
which is justified by the result we obtain, but is questionable in the first place because $k$ gets arbitrarily large values. One could derive the spectrum directly in the continuous theory via the eigenvalue problem $-\frac{d^2}{dx^2} f(x)=\lambda f(x)$, $f(0)=f(L)=0$, which obviously leads to \eqref{eq:kernel_K}. In Appendix \ref{app:discrete_calculation} we calculate the matrix elements  $\Omega_{ij}$ and $\left(\Omega^{-1}\right)_{ij}$ in the discretized theory and obtain the corresponding kernels at the very end of the calculation.}. The kernels $\Omega$ and $\Omega^{-1}$ read
\begin{align}
\Omega(x,x^\prime)&=\frac{2}{L}\sum_{k=1}^{\infty}\omega_k\sin\frac{k \pi x}{L}\,\sin\frac{k \pi x^\prime}{L},\\
\Omega^{-1}(x,x^\prime)&=\frac{2}{L}\sum_{k=1}^{\infty}\frac{1}{\omega_k}\sin\frac{k \pi x}{L}\,\sin\frac{k \pi x^\prime}{L}.
\end{align}
In the massless limit it is possible to obtain a closed form of these kernels, namely,
\begin{align}
\Omega(x,x^\prime)&=\frac{\pi}{4L^2}\left(\frac{1}{\sin^2\frac{\pi (x+x^\prime)}{2L}}-\frac{1}{\sin^2\frac{\pi (x-x^\prime)}{2L}}\right)\label{eq:kern_Omega},\\
\Omega^{-1}(x,x^\prime)&=-\frac{1}{\pi}\ln\left\vert\frac{\sin\frac{\pi\left(x-x^\prime\right)}{2L}}{\sin\frac{\pi\left(x+x^\prime\right)}{2L}}\right\vert,\label{eq:Omega_inv}
\end{align}
which coincide with the massless limit of the kernels appearing in \cite{Callan:1994py}. Notice that $\Omega(x,x^\prime)$ can be written as
\begin{equation}
\Omega(x,x^\prime)=-\frac{\pi}{L^2}\frac{\sin\frac{\pi x}{L}\sin\frac{\pi x^\prime}{L}}{\left(\cos\frac{\pi x}{L}-\cos\frac{\pi x^\prime}{L}\right)^2},
\end{equation} 
implying that this kernel is non-positive and it vanishes only when either variable is $0$ or $L$. This in turn implies that $\Omega^{-1}(x,x^\prime)$ is non-negative  and it vanishes only when either variable is $0$ or $L$.

In the continuous theory, the kernels associated to blocks of the matrices are given by the same expressions as the kernels associated to the matrices themselves, but the range of values of the corresponding coordinates is restricted appropriately. The entangling ``surface" is considered to lie between the nodes $n$ and $n+1$. In the continuous theory it lies at $x=\ell$ where  
\begin{equation}
\ell=\left(n+\frac{1}{2}\right)a.
\end{equation}
As an indicative example, $\Omega_A(x,x^\prime)$ is given by the same expression as $\Omega(x,x^\prime)$, namely equation \eqref{eq:kern_Omega}, where both $x$ and $x^\prime$ are valued in $\left[0,\ell\right]$. In a similar manner for $\Omega_B(x,x^\prime)$ it follows that $x\in\left[0,\ell\right]$ and $x^\prime\in\left[\ell,L\right]$.

At this point let us make a technical remark. So far we have formulated the calculation of entanglement entropy in terms of the matrix $M=\left(\Omega^{-1}\right)_C\Omega_C$. Given the fact that $\Omega^{-1}\Omega=I$, the blocks of $\Omega$ obey the equation $\left(\Omega^{-1}\right)_C\Omega_C+\left(\Omega^{-1}\right)_B^T\Omega_B=I_{N-n} $. As a result, we could equally well use the matrix $\tilde{M}$ defined as $\tilde{M}:=\left(\Omega^{-1}\right)_B^T\Omega_B=I_{N-n}-M$. The matrix $\Xi$ is related to the matrix $\tilde{M}$ as
\begin{equation}\label{eq:Xi_M_tilde}
\Xi=\frac{\sqrt{1-\tilde{M}}-1}{\sqrt{1-\tilde{M}}+1}.
\end{equation}
In the continuum limit the kernel $\tilde{M}(x,x^\prime)$ is given by
\begin{equation}\label{eq:kernel_M_tilde}
\tilde{M}(x,x^\prime)=\int_0^\ell dy\, \Omega^{-1}(x,y)\Omega(y,x^\prime),
\end{equation}
where both $x$ and $x^\prime$ are valued in $[\ell,L]$.  This way, we avoid the UV singularities that appear when $y\rightarrow x$ or $y\rightarrow x^\prime$. These are related to the fact that the kernels $M$ and $\tilde{M}$ obey $M(x,x^\prime)+\tilde{M}(x,x^\prime)=\delta(x-x^\prime)$. Formulating the calculation in terms of $\tilde{M}(x,x^\prime)$ allows us to avoid this delta function singularity.

The kernel \eqref{eq:kernel_M_tilde} coincides with the kernel used in \cite{Callan:1994py}. Even though it was not explicitly stated, this work derived the spectrum of the reduced density matrix for a subsystem that is half the infinite line and not just the entanglement entropy. In the next section we derive the entanglement spectrum in the case of an overall system of finite length $L$ and a subsystem corresponding to the interval $[\ell,L]$.

\section{Entanglement of (1+1)-dimensional Field Theory in a Finite Interval\protect\label{sec:Finite_Interval}}
We consider (1+1)-dimensional field theory defined in $[0,L]$. We trace out the degrees of freedom in the interval $[0,\ell]$. We would like to calculate the spectrum of the reduced density matrix and the modular Hamiltonian. As we have shown in Sections \ref{sec:Srednicki_review} and \ref{sec:Continuous_Theory}, in order to achieve this goal we have to derive the spectrum and the eigenfunctions of the kernel $\tilde{M}(x,x^\prime)$, which is defined in \eqref{eq:kernel_M_tilde}. Substituting \eqref{eq:kern_Omega} and \eqref{eq:Omega_inv} in equation \eqref{eq:kernel_M_tilde}, this kernel reads
\begin{equation}
\tilde{M}(x,x^\prime)=\frac{1}{4L^2}\int_0^\ell dy \left(\frac{1}{\sin^2\frac{\pi (x^\prime-y)}{2L}}-\frac{1}{\sin^2\frac{\pi (x^\prime+y)}{2L}}\right)\ln\frac{\sin\frac{\pi\left(x-y\right)}{2L}}{\sin\frac{\pi\left(x+y\right)}{2L}},
\end{equation}
where $x\in\left[\ell,L\right]$ and $x^\prime\in\left[\ell,L\right]$ \footnote{\label{foot:regulator} As we discussed in the previous section, using the kernel $\tilde{M}(x,x^\prime)$, instead of $M(x,x^\prime)$, allows us to avoid the singularities at $y\rightarrow x$ and at $y\rightarrow x^\prime$. However there is still one singularity left that appears when $x\rightarrow\ell$ or $x^\prime\rightarrow\ell$ and $y\rightarrow\ell$.}. The integral can be performed analytically and we obtain
\begin{multline}\label{eq:M_tilde_def}
\tilde{M}(x,x^\prime)=\frac{1}{\pi L}\frac{\cos\frac{\pi \ell}{L}-\cos\frac{\pi x}{L}}{\cos\frac{\pi x}{L}-\cos\frac{\pi x^\prime}{L}}\Bigg[\frac{\sin\frac{\pi x}{L}}{\cos\frac{\pi \ell}{L}-\cos\frac{\pi x}{L}} \ln\frac{\sin\frac{\pi\left(x^\prime-\ell\right)}{2L}}{\sin\frac{\pi\left(x^\prime+\ell\right)}{2L}} \\
-\frac{\sin\frac{\pi x^\prime}{L}}{\cos\frac{\pi\ell}{L}-\cos\frac{\pi x^\prime}{L}}\ln\frac{\sin\frac{\pi\left(x-\ell\right)}{2L}}{\sin\frac{\pi\left(x+\ell\right)}{2L}}\Bigg].
\end{multline}
The eigenvalue problem to be solved reads 
\begin{equation}\label{eq:eig_problem}
\int_\ell^L dx^\prime\,\tilde{M}(x,x^\prime)f(x^\prime)=\lambda f(x).
\end{equation}
There are continuously infinite eigenfunctions and eigenvalues. We parametrized them with the continuous variable $\omega$. They read
\begin{equation}
f(x;\omega)=\sin(\omega u(x)),\quad \lambda(\omega)= -\frac{1}{\sinh^2\left(\pi\omega\right)},\label{eq:sp}
\end{equation}
where $u(x)$ is defined as
\begin{equation}\label{eq:u_of_x}
u(x)=\ln\frac{\sin\frac{\pi\left(x+\ell\right)}{2L}}{\sin\frac{\pi\left(x-\ell\right)}{2L}}.
\end{equation}

In order to find an analytic expression for the modular Hamiltonian we need the spectral decomposition of $\tilde{M}(x,x^\prime)$. This kernel is not symmetric; Its left and right eigenfunctions do not coincide. After some algebra one obtains
\begin{equation}\label{eq:Mtilde_decomp}
\tilde{M}(x,x^\prime)=\frac{2}{L}\frac{\cosh u(x^\prime)-\cos\frac{\pi \ell}{L}}{\sin \frac{\pi \ell}{L}}\int_0^\infty d\omega \left[\frac{-1}{\sinh^2\left(\pi\omega\right)}\right]\sin(\omega u(x))\sin(\omega u(x^\prime)),
\end{equation}
where the square bracket contains the eigenvalues $\lambda(\omega)$. Obviously, this expression can be utilized in order to obtain an integral representation of any function of $\tilde{M}(x,x^\prime)$.

More details on the derivation of the spectrum and the eigenfunctions of $\tilde{M}(x,x^\prime)$ are presented in Appendix \ref{app:appendix}.

\subsection{Entanglement Entropy\protect\label{subsubsec:EE}}
In order to calculate the entanglement entropy in the continuous theory we may write equation \eqref{eq:SEE_sum} in the form
\begin{equation}
S_{\textrm{EE}}=\Tr\left[-\ln(1-\Xi)-\frac{\Xi}{1-\Xi}\ln\Xi\right].
\end{equation}
Then, using the spectral decomposition \eqref{eq:Mtilde_decomp}, we obtain
\begin{equation}\label{eq:SEE_int}
S_{\textrm{EE}}=\frac{2}{L}\int_{0}^{\infty}d\omega\, S(\omega)\int_{\ell+\epsilon}^{L} dx \frac{\cosh u(x)-\cos\frac{\pi \ell}{L}}{\sin \frac{\pi \ell}{L}}\sin^2(\omega u(x)),
\end{equation}
where $S(\omega)$ is the contribution of each frequency $\omega$ to the entanglement entropy and it is given by
\begin{equation}\label{eq:S_of_omega}
S(\omega)=\frac{2\pi\omega}{e^{2\pi\omega}-1}-\ln\left(1-e^{-2\pi\omega}\right).
\end{equation}
Since the integral over $x$ diverges, we introduced a regulator $\epsilon$, see footnote \ref{foot:regulator}. Changing the integration variable $x$ to $u$ via the inversion of equation \eqref{eq:u_of_x}, we obtain
\begin{equation}\label{eq:SEE_integrals}
S_{\textrm{EE}}=\frac{1}{\pi}\int_{0}^{\infty}d\omega \, S(\omega)\int_{0}^{u_{\textrm{max}}} du - \frac{1}{\pi}\int_{0}^{\infty}d\omega \, S(\omega)\int_{0}^{u_{\textrm{max}}}\cos(2\omega u),
\end{equation}
where
\begin{equation}
u_{\textrm{max}}=\ln\left(\frac{2L}{\pi\epsilon}\sin\frac{\pi\ell}{L}\right).
\end{equation}
Let us focus on the first term of \eqref{eq:SEE_integrals}. The integral over $u$ trivially is equal to $u_{\textrm{max}}$. The integral over $\omega$ is an integral that appears frequently in statistical physics. It is equal to $\zeta(2)/\pi=\pi/6$, where $\zeta(z)$ is Riemann's zeta function. Regarding the second term of \eqref{eq:SEE_integrals} we may send $u_{\textrm{max}}\rightarrow\infty$ so that the integral over $u$  is equal to a delta function, i.e.
\begin{equation}
\lim_{u_{\textrm{max}}\rightarrow\infty}\frac{1}{\pi}\int_{0}^{u_{\textrm{max}}} du \cos(2\omega u)= \frac{1}{2}\delta(\omega).
\end{equation}
As this function is non-vanishing only on $\omega=0$ and this is the lower bound of integration of the integral over $\omega$, there is no contribution to the entanglement entropy. In other words, one may regularize the lower bound of integration of the integral over $\omega$, and since we are interested in $\omega\rightarrow 0^+$, there is no contribution from the delta function. Putting everything together, we obtain the well known formula \cite{Holzhey:1994we}
\begin{equation}
S_{\textrm{EE}}=\frac{1}{6} \ln\left(\frac{2L}{\pi\epsilon}\sin\frac{\pi\ell}{L}\right).
\end{equation}

There is an alternative approach to this calculation that we will outline in order to make some comments. As it is evident from equation \eqref{eq:sp} and \eqref{eq:u_of_x}, the eigenfunctions automatically obey Dirichlet boundary conditions at $x=L$. The regulator $\epsilon$ at the lower bound of integration of \eqref{eq:SEE_int} can be utilized in order to impose Dirichlet boundary conditions at $x=\ell+\epsilon$. Doing so, the spectrum is discretized according to 
\begin{equation}\label{eq:discrete_spectrum}
\omega_n \ln\left(\frac{2L}{\pi\epsilon}\sin\frac{\pi\ell}{L}\right)=n\pi,\quad n\in\mathbb{N}^*
\end{equation}
Then, the entanglement entropy is given by the discrete sum appearing in equation \eqref{eq:SEE_sum}, where $\xi_i =e^{-2\pi\omega_i}$ and the upper bound of summation is set to infinity. For $L\gg \epsilon$ the eigenvalues become dense and they can be approximated  by a continuous distribution \cite{Callan:1994py}. In this approach it is evident that entanglement entropy is divergent because the density of eigenvalues blows up. The divergence is associated to neither the UV nor the IR of the theory. However, this line of reasoning is linked to a fine technical detail; in order to properly perform the above calculation one has to implement the Euler-McLaurin summation formula. The leading integral term yields the above result, but there are also remainder terms. Due to the fact that \eqref{eq:SEE_sum} is singular at $\xi_i=1$ the latter include divergent terms. These terms should be neglected, but in this approach doing so is kind of ad hoc. In our approach this fact is more transparent.
\subsection{Modular Hamiltonian}
Having calculated the entanglement entropy, in this section we derive the corresponding modular Hamiltonian. Equation \eqref{eq:modular_disc} is the modular Hamiltonian of the discrete theory. In the continuous theory the modular Hamiltonian is given by
\begin{equation}\label{eq:modular_kernel}
\hat{H}_M=\int dx \int dy\,\left(\Pi(x)\tilde{K}\left(x,y\right)\Pi(y)+\Phi(x)\tilde{V}\left(x,y\right)\Phi(y)\right),
\end{equation}
where $\Pi(x)$ is the conjugate momentum of the field $\Phi(x)$, i.e. it is the operator $-i \frac{\delta}{\delta \Phi(x)}$. The kernels $\tilde{K}\left(x,y\right)$ and $\tilde{V}\left(x,y\right)$ are the continuum limit of the matrices defined in equations \eqref{eq:modular_K} and \eqref{eq:modular_V}, respectively.

Using equation \eqref{eq:Omega_inv} along with \eqref{eq:Mtilde_decomp}, it turns out that the kernel $\tilde{K}(x,y)$ is given by
\begin{equation}\label{eq:tilde_K_init}
\tilde{K}(x,y)=-\frac{2}{\pi}\int_0^\infty du^\prime \ln\left\vert\frac{\sinh\frac{u^\prime-u_y}{2}}{\sinh\frac{u^\prime+u_y}{2}}\right\vert \int_0^\infty d\omega\, \omega\tanh\left(\pi\omega\right)\sin(\omega u_x)\sin(\omega u^\prime),
\end{equation}
where $x\in\left(\ell, L\right]$ and $y\in\left(\ell, L\right]$. We also used the notation $u_x\equiv u(x)$, $u_y\equiv u(y)$ and $u^\prime\equiv u(x^\prime)$. The function $u(x)$ is defined in equation \eqref{eq:u_of_x}. After some algebra that is presented in Appendix \ref{app:appendix}, one obtains
\begin{equation}\label{eq:kernel_K_final_u}
\tilde{K}(x,y)=\pi\delta\left(u_x-u_y\right).
\end{equation}
In terms of the original coordinates the kernel reads
\begin{equation}\label{eq:kernel_K_final}
\tilde{K}(x,y)= L\frac{\cos\frac{\pi \ell}{L}-\cos\frac{\pi x}{L}}{\sin\frac{\pi \ell}{L}}\delta\left(x-y\right).
\end{equation} 

Similarly, the kernel $\tilde{V}(x,y)$ is given by the integral
\begin{multline}\label{eq:tilde_V_init}
\tilde{V}(x,y)=\frac{\pi}{2L^2}\frac{\left(\cosh u_x-\cos\frac{\pi \ell}{L}\right)\left(\cosh u_y-\cos\frac{\pi \ell}{L}\right)}{\sin^2\frac{\pi \ell}{L}}\int_0^\infty d\omega \frac{\omega\sin(\omega u_y)}{\coth\left(\pi\omega\right)} \\ \times\int_0^\infty du^\prime \sin(\omega u^\prime) \left[\frac{1}{\sinh^2\frac{u_x+u^\prime}{2}}-\frac{1}{\sinh^2\frac{u_x-u^\prime}{2}}\right].
\end{multline}
After some algebra that is also presented in Appendix \ref{app:appendix}, we may write this formula as
\begin{equation}\label{eq:kernel_V_final_u}
\tilde{V}(x,y)=-\pi\frac{d u_x}{dx}\frac{d u_y}{dy}\frac{d^2}{d u_x^2}\delta\left(u_x-u_y\right).
\end{equation}
In terms of the original coordinates the kernel reads
\begin{equation}\label{eq:kernel_V_final}
\tilde{V}(x,y)= -L\frac{d}{dx}\left[\frac{\cos\frac{\pi \ell}{L}-\cos\frac{\pi x}{L}}{\sin\frac{\pi \ell}{L}}\frac{d}{dx}\delta\left(x-y\right)\right].
\end{equation}

As expected for this specific setup \cite{Cardy:2016fqc}, the modular Hamiltonian is a local functional of the stress-energy tensor. Substituting our results in \eqref{eq:modular_kernel} we obtain
\begin{equation}\label{eq:Modular_Large}
\hat{H}_M=2\pi\frac{L}{\pi}\int_{\ell}^{L} dx \frac{\cos\frac{\pi \ell}{L}-\cos\frac{\pi x}{L}}{\sin\frac{\pi \ell}{L}}\,T_{00}(x),
\end{equation}
where we dropped a surface term and $T_{00}(x)$ reads
\begin{equation}
T_{00}(x)=\frac{1}{2}\Pi^2(x)+\frac{1}{2}\left(\partial_x\Phi\left(x\right)\right)^2.
\end{equation}
One may consider the complementary subsystem, i.e. the degrees of freedom in $\left[0,\ell\right]$. Then the above formula assumes the form
\begin{equation}\label{eq:Modular_Small}
\hat{H}_M=2\pi\frac{L}{\pi}\int_{0}^{\ell} dx \frac{\cos\frac{\pi x}{L}-\cos\frac{\pi \ell}{L}}{\sin\frac{\pi \ell}{L}}\,T_{00}(x).
\end{equation}

For a theory defined on an infinite half-line \cite{Casini:2011kv,Cardy:2016fqc}, i.e. $L\gg \ell$, using \eqref{eq:Modular_Small} we obtain
\begin{equation}\label{eq:Modular_dS}
\hat{H}_M=2\pi\int_{0}^{\ell} dx \frac{\ell^2-x^2}{2\ell}\,T_{00}(x).
\end{equation}
Finally, one may study the case of a theory defined on an infinite line and a subsystem consisting of the degrees of freedom in the region $x>0$ \cite{Bisognano:1975ih,Bisognano:1976za}. This is achieved by defining the theory in $[-L/2,L/2]$, setting $\ell=0$ and considering the limit $L\gg x$. Doing so, one obtains
\begin{equation}
\hat{H}_M=2\pi\int_{0}^{\infty} dx\, x\,T_{00}(x).
\end{equation}
\section{Discussion\protect\label{sec:discussion}}

A quantum system is always quantum even if it appears to have classical behaviour. In this context it is natural to wonder how classical gravity emerges from a quantum theory. The fact that a bottom-up approach, i.e. the quantization of general relativity, is plagued with dozens of problems indicates that we need new principles. The AdS/CFT correspondence suggest that entanglement, which is the very essence of quantum mechanics in the sense that it is a property with no classical analogue, may serve as a source of gravity, which is to be interpreted as an emergent entropic force  \cite{VanRaamsdonk:2010pw,VanRaamsdonk:2009ar,Lashkari:2013koa,Faulkner:2013ica}. If we really believe that gravity and quantum mechanics are essentially the same \cite{Susskind:2017ney}, we should be able to support this claim beyond the framework of AdS/CFT.

In this quest, we have to relate the dynamics of a subsystem, which is the dynamics perceived by an observer who does not have access to observables of the complementary subsystem, to some kind of gravitational dynamics. This information is encoded in the modular Hamiltonian $H_A$. In order to derive $H_A$, one needs not only the whole entanglement spectrum, but also the corresponding eigenfunctions of the reduced density matrix. For this reason we demonstrate how the modular Hamiltonian can be derived via the continuum limit of Srednicki's method \cite{Srednicki:1993im}, which is the appropriate method for this purpose.

In the continuum limit the modular Hamiltonian is given by \eqref{eq:modular_kernel}. In section \eqref{sec:Mudular_2d} we obtained explicit expressions for the kernels appearing in \eqref{eq:modular_kernel} in terms of the original coordinates $x$ and $y$. Let us express it in terms of the coordinates $u_x=u(x)$ and $u_y=u(y)$, defined via the coordinate transformation \eqref{eq:u_of_x}. Substituting the kernels \eqref{eq:kernel_K_final_u} and \eqref{eq:kernel_V_final_u} we obtain
\begin{equation}
\hat{H}_M=2\pi\int_0^\infty du_x \frac{1}{2}\left[\Pi(u_x)^2-\Phi(u_x)\frac{d^2}{d u_x^2}\Phi(u_x)\right],
\end{equation}
where we used the fact that $\Pi(x)=\frac{d u_x}{dx}\Pi(u_x)$, which is required in order to preserve the canonical commutation relations. In these coordinates the modular Hamiltonian density coincides with the Hamiltonian density of the overall theory. The reduced density matrix describes a system at finite temperature $2\pi$. It follows that the mapping \eqref{eq:u_of_x} actually maps the interval $(\ell,L]$ to the Rindler wedge\footnote{The same mapping is used in \cite{Cardy:2016fqc}.}. The Unruh effect \cite{Unruh:1976db} is imprinted in the modular Hamiltonian. Regarding infinite-size systems, works from the early 90s have argued about this fact based on the interpretation of the Euclidean path integral \cite{Kabat:1994vj,Dowker:1994fi}. In this work we reach the same conclusion relying only on elementary quantum mechanics. For an observer that performs local measurements inside a subsystem, the density matrix that determines the outcome of all measurements is thermal\footnote{For the discrete system each mode of the subsystem lies in a thermal state, but the corresponding temperatures differ \cite{Srednicki:1993im}.}. The subsystem appears at finite temperature and there is no local measurement that could indicate otherwise. The physical reality of this observer is emergent. 

It is well known that entanglement entropy at the ground state of free scalar field theory obeys an area law \cite{Srednicki:1993im,Bombelli:1986rw}. The similarity between the entropy of black holes and entanglement entropy is very intriguing, but, as we discussed, we would like to take things further. In this spirit we revisited some classical works on entanglement in quantum field theory. We analyzed a system of coupled harmonic oscillators and showed that Srednicki's method, the replica trick and the correlation functions method for the calculation of entanglement entropy are interrelated and obviously equivalent. We applied this formalism to the discretized $(1+1)-$dimensional free scalar field theory, defined on a finite interval. Working in the framework of quantum mechanics, all formulae are well-defined and there are no ambiguities related to the definition of traces, the measure of path integrals etc. Having set up a well-posed problem, we studied its continuum limit. We related the eigenvalues of the reduced density matrix to the eigenvalues of a bilinear kernel. We managed to solve this problem even when the overall system has finite length. Thus, we obtained the entanglement spectrum of $(1+1)-$dimensional scalar field theory defined on a finite interval, which is a novel result.

In this work we developed a new technique and applied it in the well studied $(1+1)$-dimensional free scalar field theory. Nevertheless this methodology can be utilized in many other systems. 

% shortcomings

In the quest to probe further the relation between entanglement and gravity, which is a major motivation of this work, a basic shortcoming is the fact that most calculations of entanglement entropy concern free field theories. In these theories an entangling surface, i.e. a physical obstacle preventing the measurements in a specific region, can not be created dynamically. In an interacting strongly-coupled theory the presence of a non-perturbative object prevents particle excitations to probe its region. This mechanism introduces an effective ``Schwarzschild radius" that relates the radius of the entangling surface to physical parameters for observers who use apparatuses that measure the fundamental excitations of the field. In particular the radius of the entangling surface would be related to the energy of the probe particle. On the contrary, in free field theory the radius of the entangling surface is set by hand. It would be very interesting to study entanglement entropy and the modular Hamiltonian in such setups.

Besides probing further the relation between gravity and quantum mechanics, as we did by identifying the Unruh effect in the modular Hamiltonian, it would be interesting to employ this interplay between the discretized and continuous theory in order to study different aspects of entanglement entropy in field theory. For instance, one could study spherical entangling surfaces in higher dimensions, by  expressing the problem in terms of effective $(1+1)$-dimensional radial systems, one for each angular momentum sector\footnote{One could also implement quantum graphs \cite{Giavoni:2023cwg}. }. Such an approach is followed in \cite{Huerta:2022tpq}, but the starting point of this work is the modular Hamiltonian and not the discretized theory. An equally interesting problem would be the study of non-compact subsystems in $(1+1)$-dimensional field theory, interacting theories and free theories on curved backgrounds \cite{Boutivas:2024sat}.

The techniques we developed could also be used for the study of other measures of entanglement and related quantities such as the entanglement negativity \cite{Calabrese:2012ew}, the capacity of entanglement \cite{Yao_2010}, the symmetry resolved entanglement \cite{Pirmoradian:2023uvt} and the reflected entropy \cite{Bueno:2020fle}.

 %%%%%%%%%%%%%%%%%%%%%%%%%%%%%%%%%%%%%%%%%%%%%%%%%%%%%%%%%%%%%%%%%%%%%%%%%%%%

Another aspect of this work regards the formal definition of entanglement entropy. Entanglement entropy in QFT is divergent and is considered ill-defined. This divergence originates neither from the UV nor the IR. It originates from the density of states of the subsystem, which diverges  in the continuous theory \cite{Callan:1994py}. The origin of this divergence is related to the question of whether the Hilbert space $\mathcal{H}$ of the overall system $A\cup  A^C$ factorizes as $\mathcal{H}_A\times \mathcal{H}_{A^C}$ \footnote{Even in the case of lattice gauge theory, because of Gauss's law, the Hilbert spaces do not factorize. One has to consider a larger Hilbert space, which relaxes the constraints imposed by the Gauss's law at the entangling surface \cite{Buividovich:2008gq,Donnelly:2011hn,Casini:2013rba,Ghosh:2015iwa}.}.

On a more formal basis, entanglement entropy is associated with the algebra of observables corresponding to the subsystem $A$. This is a von Neumann algebra and depending on its type, one may be unable to even define a trace.  This subject has drawn a lot of attention recently \cite{Witten:2018zxz,Witten:2018zva,Witten:2021jzq,Longo:2022lod,Chandrasekaran:2022eqq,Witten:2022xxp,Witten:2023qsv,Kudler-Flam:2023qfl}, see also \cite{Jafferis:2015del,Dong:2016eik,Cotler:2017erl}. This problem is usually approached in the infinite $N$ limit. The behavior of the divergences is improved by  $1/N$ corrections  \cite{Leutheusser:2021qhd,Leutheusser:2021frk}.

In this work we used lattice discretization, resulting in a quantum mechanical system, where entanglement entropy is well-defined and we took the continuum limit after solving the eigenvalue problem of the reduced density matrix. Taking the continuum limit at the very end of the calculation would be an interesting improvement of our work regarding this issue.

\paragraph{Acknowledgments}
Part of the research of D.K. was supported by the FAPESP Grant No. 2021/01819-0.
\appendix
\section{Algebraic Construction of the Eigenstates of the Reduced Density Matrix\protect\label{sec:Algebraic}}
In this section we derive the spectrum and the eigenstates of the reduced density matrix, which were obtained in Section \ref{subsec:eigenstates_main}, using creation and annihilation operators.
\subsection{The Ground Eigenfunction}
The reduced density matrix given by equation \eqref{eq:rho_red_g_b} has a Gaussian ``ground" eigenfunction of the form
\begin{equation}
\Psi_0 \left( x \right) =c_0 \exp \left( - \frac{1}{2} x^T \mathcal{A}\,x \right).
\end{equation}
It is straightforward to show that
\begin{equation}
\int_{-\infty}^{\infty} dx^\prime \rho_\textrm{red}(x,x^\prime)\Psi_0 \left( x^\prime \right)= \sqrt{ \frac{2^{N-n}\det \left( \gamma - \beta \right)}{\det \left(\gamma+\mathcal{A}\right)} }\Psi_0 \left( x \right),
\end{equation}
provided that $\mathcal{A}$ satisfies the equation
\begin{equation}
\mathcal{A}=\gamma-\beta\left(\gamma+\mathcal{A}\right)^{-1}\beta.
\end{equation}
The solution of this equation that corresponds to a normalizable eigenfunction is
\begin{equation}\label{eq:matA_def}
\mathcal{A}=\gamma \sqrt{I-\left(\gamma^{-1}\beta\right)^2}=\gamma^{1/2} \sqrt{I-\hat{\beta}^2}\gamma^{1/2},
\end{equation}
where $\hat{\beta}$ is given by \eqref{eq:beta_hat}. It is a matter of algebra to show that
\begin{equation}
\left( \frac{2^{N-n}\det \left( \gamma - \beta \right)}{\det \left(\gamma+\mathcal{A}\right)} \right)^{\frac{1}{2}}=\det\left(I-\Xi\right),
\end{equation}
where $\Xi$ is defined in \eqref{eq:definition_Xi}, i.e. the eigenvalue corresponding to the ``ground" eigenfunction is equal to $\det\left(I-\Xi\right)$.

\subsection{The First Excited Eigenfunctions}
Having obtained the ``ground" eigenfunction, we search for eigenfunctions of the form
\begin{equation}
\Psi_{1i}\left(x\right)=c_{1i} \left(v_i^T x\right) \exp\left(-\frac{1}{2} x^{T}\mathcal{A}\,x\right),
\end{equation}
where $v_i$ is a column matrix and $i=1,\dots ,N-n$. It is a matter of algebra to show that
\begin{equation}
\int_{-\infty}^{\infty} dx^\prime \rho_\textrm{red}(x,x^\prime)\Psi_{1i} \left( x^\prime \right) = c_{1i} \det\left(I-\Xi\right) \left(v_i^T\Xi\, x\right)\exp\left(-\frac{1}{2} x^{T}\mathcal{A}\,x\right),
\end{equation}
Thus, $\Psi_{1i}\left(x\right)$ is an eigenfunction provided
\begin{equation}
\Xi^T v_i=\xi_i v_i.
\end{equation}
Thus, $v_i$ is a left eigenvector of the matrix $\Xi$ and $\xi_i$ is the corresponding eigenvalue. The eigenvalues of the reduced density matrix read $\det\left(I-\Xi\right)\xi_i$. Since $\Xi$ is a function of the matrix $\gamma^{-1}\beta$, see equation \eqref{eq:definition_Xi}, this requirement is equivalent to
\begin{equation}\label{eq:xi_i_def}
\beta\gamma^{-1}v_i=\hat{\beta}_iv_i,\quad \xi_i=\frac{\hat{\beta}_i}{1+\sqrt{1-\hat{\beta}_i^2}}.
\end{equation}
Moreover, the above relation implies that
\begin{equation}
\hat{\beta}\left(\gamma^{-1/2}v_i\right)=\hat{\beta}_i\left(\gamma^{-1/2}v_i\right).
\end{equation}
Since $\gamma^{-1/2}v_i$ are the eigenvectors of a Hermitian matrix, they can be normalized, so that
\begin{equation}
v_i^T\gamma^{-1}v_j=\delta_{ij}.
\end{equation}

In the following we use component notation for the eigenvectors $v_i$, which follows the convention $v_j^Tx=v^j_{k}x_k$. In particular, we are interested in the normalization and completeness relations
\begin{equation}\label{eq:eig_normalization}
v^i_a\gamma^{-1}_{ab}v_b^j=\delta_{ij},\quad \delta_{ab}=\sum_{i=1}^{N-n}v_a^iv_c^i\gamma^{-1}_{cb},
\end{equation}
as well as in the fact that \eqref{eq:xi_i_def} assumes the form
\begin{equation}\label{eq:beta_decomposition}
\left(\beta\gamma^{-1}\right)_{ab}=\sum_{i=1}^{N-n}\hat{\beta}_i v_a^iv_c^i\gamma^{-1}_{cb}.
\end{equation}

In order to complete the derivation we have to specify the normalization constant $c_{1i}$. To do so, we calculate the inner product of the eigenfunctions $\Psi_{1i}$ and $\Psi_{1j}$. It is straightforward to show that
\begin{equation}\label{eq:normalization_Psi1}
\int d x\, \Psi_{1i}\left(x\right)\Psi^*_{1j}\left(x\right) =c^{\phantom{*}}_{1i}c_{1j}^*\int dx \left(v_i^Tx\right)\left(v_j^Tx\right)\exp\left(-x^{T} \mathcal{A}\,x \right)=\frac{c^{\phantom{*}}_{1i}c_{1j}^*}{2c_0^2}v^T_j\mathcal{A}^{-1}v_i.
\end{equation}
Recalling that $\mathcal{A}=\gamma\sqrt{1-\left(\gamma^{-1}\beta\right)^2}=\sqrt{1-\left(\beta\gamma^{-1}\right)^2}\gamma$, it follows that the matrix $\mathcal{A}^{-1}$ is given by the expression $\mathcal{A}^{-1}=\gamma^{-1}\left(1-\left(\beta\gamma^{-1}\right)^2\right)^{-1/2}$. As a result, the right-hand-side of \eqref{eq:normalization_Psi1} is given by
\begin{equation}
\frac{c^{\phantom{*}}_{1i}c_{1j}^*}{2c_0^2}\mathbf{v}^T_j\mathcal{A}^{-1}\mathbf{v}_i=\frac{\vert c_{1i}\vert^2}{2c_0^2\alpha_i}\delta_{ij},
\end{equation}
where $\alpha_i=\sqrt{1-\hat{\beta}^2_i}$. Thus, the normalization constant of the eigenstates $\Psi_{1i}(x)$ reads
\begin{equation}
c_{1i}=\sqrt{2\alpha_i}c_0.
\end{equation}
\subsection{Creation and Annihilation Operators}
Using the ``ground" and the ``first excited" eigenfunctions, we may introduce creation and annihilation operators that connect them. In the following when an index variable appears twice in a single term, summation of that term over all the values of the index is implied. We impose that the annihilation operators $\hat{A}_i$ have the form
\begin{equation}
\hat{A}_i=C_{ik}\left(\frac{\partial}{\partial x_k}+\mathcal{A}_{k\ell}x_\ell\right),
\end{equation}
where $\mathcal{A}$ is given by \eqref{eq:matA_def}. It is straightforward to show that $\hat{A}_i \Psi_0\left(x\right)=0$. Furthermore, acting with the annihilation operator $\hat{A}_i$ on $\Psi_{1j}\left(x\right)$ yields
\begin{equation}
\hat{A}_i \Psi_{1j}\left(x\right)=\sqrt{2\alpha_i} C_{ik}v^j_{k}\Psi_{0}\left(x\right).
\end{equation}
Taking equation \eqref{eq:eig_normalization} into account, we obtain that $\hat{A}_i \Psi_{1j}\left(x\right)=\delta_{ij}\Psi_{0}\left(x\right)$ as long as
\begin{equation}
C_{ik}=\frac{1}{\sqrt{2\alpha_i}}v^{i}_{\ell}\left(\gamma^{-1}\right)_{\ell k}.
\end{equation}
Putting everything together, the annihilation and creation operators read
\begin{align}
\hat{A}_i&=\frac{v^{i}_{m}}{\sqrt{2\alpha_i}}\left(\alpha_i x_m+\left(\gamma^{-1}\right)_{m k}\frac{\partial}{\partial x_k}\right),\label{eq:op_A}\\
\hat{A}_i^\dagger&=\frac{v^{i}_{m}}{\sqrt{2\alpha_i}}\left(\alpha_i x_m-\left(\gamma^{-1}\right)_{m k}\frac{\partial}{\partial x_k}\right).\label{eq:op_A_dag}
\end{align}

It is straightforward to show that $\left[\hat{A}_i,\hat{A}_j^\dagger\right]=\delta_{ij}$ and $\left[\hat{A}_i,\hat{A}_j\right]=0$. For what follows, it is helpful to invert these relations and obtain
\begin{align}
x_k=&\sum_{i=1}^{N-n}\frac{1}{\sqrt{2\alpha_i}}\left(\gamma^{-1}\right)_{kn}v^{i}_{n}\left(\hat{A}_i+\hat{A}_i^\dagger\right),\\
p_k=&\sum_{i=1}^{N-n}\sqrt{\frac{\alpha_i}{2}}v^{i}_{k}\left(\hat{A}_i-\hat{A}_i^\dagger\right).
\end{align}
It remains to show that these operators generate the whole tower of eigenstates of the reduced density matrix.
\subsection{The Tower of Eigenstates}
Assume that $\Psi(x)$ is an eigenfunction of the reduced density matrix $\rho_{\textrm{red}}$, i.e.
\begin{equation}\label{eq:induction_step_0}
\int dx^\prime \rho_{\textrm{red}}\left(x,x^\prime\right)\Psi\left(x^\prime\right)=\lambda\Psi\left(x\right).
\end{equation}
We will show that $\hat{A}^\dagger_i\Psi(x)$ is also an eigenfunction. For this purpose we are going to calculate the integral
\begin{equation}
\int dx^\prime \rho_{\textrm{red}}\left(x,x^\prime\right)\hat{A}^\dagger_i\Psi\left(x^\prime\right) =\frac{v^{i}_{m}}{\sqrt{2\alpha_i}}\int dx^\prime \rho_{\textrm{red}}\left(x,x^\prime\right)\Bigg[\alpha_i x_m^\prime -\left(\gamma^{-1}\right)_{m k}\frac{\partial}{\partial x^\prime_k}\Bigg]\Psi\left(x^\prime\right).
\end{equation}
Integrating by parts yields
\begin{equation}
\int dx^\prime \rho_{\textrm{red}}\left(x,x^\prime\right)\hat{A}^\dagger_i\Psi\left(x^\prime\right)=\frac{\lambda v^{i}_{m}}{\sqrt{2\alpha_i}}\bigg[\hat{\beta}_i x_m\Psi\left(x\right) -\frac{1-\alpha_i}{\lambda}\int dx^\prime \rho_{\textrm{red}}\left(x,x^\prime\right)x_m^\prime\Psi\left(x^\prime\right)\bigg].
\end{equation}
Notice that by differentiating \eqref{eq:induction_step_0} with respect to $x$ one can show that
\begin{equation}
v^{i}_{m}\int dx^\prime \rho_{\textrm{red}}\left(x,x^\prime\right)x_m^\prime\Psi\left(x^\prime\right) =\frac{\lambda v^{i}_{m}}{\hat{\beta}_i} \left[x_m+\left(\gamma^{-1}\right)_{mk} \frac{\partial}{\partial x_k}\right]\Psi\left(x\right).
\end{equation}
After some algebra we obtain
\begin{equation}
\int dx^\prime \rho_{\textrm{red}}\left(x,x^\prime\right)\hat{A}^\dagger_i\Psi\left(x^\prime\right)=\frac{\lambda\xi_iv^{i}_{m}}{\sqrt{2\alpha_i}}\left[\alpha_i x_m-\left(\gamma^{-1}\right)_{mk} \frac{\partial}{\partial x_k}\right]\Psi\left(x\right)=\lambda\xi_i\hat{A}^\dagger_i\Psi\left(x\right),
\end{equation}
where $\xi_i$ is defined in \eqref{eq:xi_i_def}. Therefore we proved that if $\Psi(x)$ is an eigenfunction of the reduced density matrix corresponding to the eigenvalue $\lambda$, then $\hat{A}^\dagger_i\Psi(x)$ is also an eigenfunction, corresponding to the eigenvalue $\lambda \xi_i$. It follows that the reduced density matrix has eigenfunctions and eigenvalues that read
\begin{align}
\Psi_{\vec{n}}(x)&=\left[\prod_{i=1}^{N-n}\frac{\left(\hat{A}^\dagger_i\right)^{n_i}}{\sqrt{n_i!}}\right]\Psi_{0}(x),\\
\lambda_{\vec{n}}&=\prod_{i=1}^{N-n}\left(1-\xi_i\right)\xi_i^{n_i}.
\end{align}
The vector $\vec{n}$ denotes the set of the non-negative integers $n_i$. We also took into account that the eigenvalue corresponding to the ``ground" eigenfunction is $\det\left(I-\Xi\right)=\prod_{i=1}^{N-n}(1-\xi_i)$. The spectrum is normalized correctly, since $\sum_{n_i=0}^\infty\left(1-\xi_i\right)\xi_i^{n_i}=1$. Essentially the reduced density matrix factorizes to $N-n$ density matrices, one for each pair of creation and annihilation operators. Each of these density matrices has normalized spectrum. Thus, there are no other eigenfunctions corresponding to non-vanishing eigenvalues.
\section{Correlation Functions of $(1+1)$-dimensional Discretized Field Theory\protect\label{app:discrete_calculation}}
In this appendix we calculate the correlation functions in $(1+1)$-dimensional discretized field theory and derive the continuum limit at the very end of the calculation resulting in the equations \eqref{eq:kern_Omega} and \eqref{eq:Omega_inv}.
\subsection{The Powers of the Coupling Matrix}
We write the coupling matrix  \eqref{eq:matK_2d} as
\begin{equation}
K_{ij}=\frac{1}{a^2}\left[k_d\delta_{i,j}-(\delta_{i,j+1}+\delta_{i+1,j})\right],
\end{equation}
where
\begin{equation}
k_d=2+\mu^2a^2.
\end{equation}
The matrix $K$ can be diagonalized as $K=O^T K_D O$, where
\begin{equation}
\left(K_D\right)_{ij}=d_k^2\delta_{i,j},\quad d_k=\frac{1}{a}\sqrt{k_d-2\cos\frac{\pi k}{N+1}}
\end{equation}
and
\begin{equation}
O_{ij}=\sqrt{\frac{2}{N+1}}\sin\frac{i j \pi}{N+1}.
\end{equation}
Obviously the half-integer powers of $K$ are given by
\begin{equation}
\Omega_{ij}^{2n+1}=\sum_{k=1}^N d_k^{2n+1} O_{ik}O_{jk}.
\end{equation}
As a result, we obtain
\begin{equation}
\Omega^{2n+1}_{ij}=\frac{1}{N+1}\sum_{k=1}^Nd^{2n+1}_k\left(\cos\frac{ \left(i-j\right)k\pi}{N+1}-\cos\frac{ \left(i+j\right)k\pi}{N+1}\right).
\end{equation}
We will use the expansions
\begin{equation}
\left(1-x\right)^{n+\frac{1}{2}}=\sum_{\ell=0}^\infty(-1)^\ell\binom{n+1/2}{\ell}x^\ell\equiv\sum_{\ell=0}^{\infty} c^{(n)+}_\ell x^\ell
\end{equation}
and
\begin{equation}
\frac{1}{\left(1-x\right)^{n+\frac{1}{2}}}=\sum_{\ell=0}^\infty\binom{n+\ell-1/2}{\ell}x^\ell\equiv\sum_{\ell=0}^{\infty} c^{(n)-}_\ell x^\ell.
\end{equation}
Using them in order to expand $d_k^{2n+1}$ we obtain
\begin{multline}
\Omega_{ij}^{\pm(2n+1)}=\frac{1}{N+1}\left(\frac{k_d^{n+\frac{1}{2}}}{a^{2n+1}}\right)^{\pm1}\sum_{\ell=0}^\infty  c^{(n)\pm}_\ell\left(\frac{2}{k_d}\right)^\ell \\ \times\sum_{k=1}^N\left(\cos\frac{ \left(i-j\right)k\pi}{N+1}-\cos\frac{ \left(i+j\right)k\pi}{N+1}\right)\cos^\ell\frac{\pi k}{N+1}.
\end{multline}
At this point we make use of the sum
\begin{equation}
\begin{split}
S(a)&=\frac{2^\ell}{N+1}\sum_{k=1}^N\cos\frac{ a k\pi}{N+1}\cos^\ell\frac{ k\pi}{N+1}\\
&=\frac{1+(-1)^{\ell+a}}{2}\Bigg[\frac{2^\ell}{N+1}+\sum_{p=-\infty}^\infty\binom{\ell}{\frac{\ell+a}{2}-p\left(N+1\right)}\Bigg],
\end{split}
\end{equation}
in order to replace the summation over $k$ with an infinite sum over finite-$N$ wrapping effects, namely
\begin{multline}
\Omega^{\pm(2n+1)}_{ij}=\left(\frac{k_d^{n+\frac{1}{2}}}{a^{2n+1}}\right)^{\pm1}\sum_{\ell=0}^\infty \frac{c^{(n)\pm}_\ell}{k_d^\ell}\frac{1+(-1)^{\ell+i+j}}{2}\\ \times\sum_{p=-\infty}^\infty\left[\binom{\ell}{\frac{\ell+i-j}{2}-p\left(N+1\right)}-\binom{\ell}{\frac{\ell+i+j}{2}-p\left(N+1\right)}\right].
\end{multline}
We interchange the summation over wrappings and the summation over powers of $k_d$ and perform the latter using the formula
\begin{equation}
\sum_{\ell=0}^\infty \frac{c^{(n)\pm}_\ell}{k_d^\ell}\frac{1+(-1)^{\ell+\alpha}}{2}\binom{\ell}{\frac{\ell+\alpha}{2}} =\frac{1}{k_d^{\vert \alpha\vert}}\binom{\alpha_\pm-1}{\vert \alpha\vert}\, _2F_1\left(\frac{\alpha_\pm}{2},\frac{\alpha_\pm+1}{2};\vert \alpha\vert +1;\frac{4}{k_d^2}\right),
\end{equation}
where $\alpha_\pm=\mp n+\vert \alpha\vert\mp\frac{1}{2}$. For the specific values of the parameters, the hypergeometric function can be expressed in terms of Associated Legendre functions of the second kind of type 3 using the formula
\begin{equation}
\, _2F_1\left(a,a+\frac{1}{2};c;z\right) =\frac{2^{c-\frac{1}{2}} \Gamma (c)}{\sqrt{\pi } \Gamma (2 a) e^{i \pi  \left(2 a-c+\frac{1}{2}\right)}}\frac{(1-z)^{\frac{1}{4} (2c-1)-a}}{z^{\frac{1}{4} (2 c-1)}}\mathfrak{Q}_{c-\frac{3}{2}}^{2a-c+\frac{1}{2}}\left(\frac{1}{\sqrt{z}}\right).
\end{equation}
Putting everything together, we obtain
\begin{align}
\Omega_{ij}^{2n+1}&=\frac{e^{(n+1) i \pi}\left(k_d^2-4\right)^{(n+1)/2}}{a^{2n+1}\sqrt{\pi}\Gamma\left(-n-\frac{1}{2}\right)}\sum_{p=-\infty}^\infty\left[\mathfrak{Q}_{\vert a_-\vert-\frac{1}{2}}^{-1-n}\left(k_d/2\right)-\mathfrak{Q}_{\vert a_+\vert-\frac{1}{2}}^{-1-n}\left(k_d/2\right)\right], \label{eq:Omega_pos}\\
\Omega^{-(2n+1)}_{ij}&=\frac{a^{2n+1}e^{-n i \pi}\left(k_d^2-4\right)^{-n/2}}{\sqrt{\pi}\Gamma\left(n+\frac{1}{2}\right)} \sum_{p=-\infty}^\infty\left[\mathfrak{Q}_{\vert a_-\vert-\frac{1}{2}}^n\left(k_d/2\right)-\mathfrak{Q}_{\vert a_+\vert-\frac{1}{2}}^n\left(k_d/2\right)\right]\label{eq:Omega_neg},
\end{align}
where $a_\pm=i\pm j-2p\left(N+1\right)$. The parameter $n$ can be analytically continued to $\mathbb{C}$ excluding the poles of the Gamma function that appear above.

\subsection{The Continuum Limit}
Unfortunately, we are unable to continue the calculation of the entanglement spectrum in the lattice model. Therefore we derive the continuum limit of \eqref{eq:Omega_pos} and \eqref{eq:Omega_neg} in order to continue the calculation. For matrices this limit is given by equation \eqref{eq:kernel_limit}. For the matrices of interest the continuum limit is obtained using the limit representation of the modified Bessel function of the 2nd kind $K$ in terms of the Associated Legendre functions of the second kind of type 3, namely
\begin{equation}
\lim_{\nu\rightarrow\infty} \nu^{-\mu}\mathfrak{Q}^\mu_\nu\left(\cosh\frac{z}{\nu}\right)=e^{\mu\pi i} K_\mu\left(z\right).
\end{equation}
For a $(1+1)-$dimensional field theory $d_k=2+\mu^2a^2$, so that 
\begin{equation}
\cosh\left(\mu a\right)=\frac{d_k}{2}+\mathcal{O}\left(a^4\right).
\end{equation}
It follows that
\begin{equation}
\begin{split}
&\lim_{a\rightarrow0^+}\frac{\left(d_k-2\right)^{(n+1)/2}}{a^{n+1}}\mathfrak{Q}_{k-\frac{1}{2}}^{-1-n}\left(d_k/2\right)\\
&\qquad=\lim_{\nu\rightarrow\infty} \left(\frac{\mu}{z^\prime}\right)^{n+1}\nu^{n+1}\mathfrak{Q}^{-1-n}_\nu\left(\cosh\frac{\mu z^\prime}{\nu}\right)\\
&\qquad=\left(\frac{\mu}{z}\right)^{n+1}\left(-1\right)^{n+1} K_{n+1}\left(\mu z\right),
\end{split}
\end{equation}
where $\nu=z^\prime/a$, $z=k a$, and $z^\prime=z-(a/2)$. We took into account that $K_{-n}\left(z\right)=K_{n}\left(z\right)$. Notice that $z$ it kept fixed as $a\rightarrow 0$. Thus, we conclude that the kernels $\Omega^{2n+1}(x,y)$ and $\Omega^{-(2n+1)}(x,x^\prime)$ are given by
\begin{align}
\Omega^{2n+1}(x,x^\prime)&=\frac{\left(2\mu\right)^{n+1}}{\sqrt{\pi}\Gamma\left(-n-\frac{1}{2}\right)}\sum_{p=-\infty}^\infty\left[\frac{K_{n+1}\left(\mu\vert x_-\vert\right)}{\vert x_-\vert^{n+1}}-\frac{K_{n+1}\left(\mu\vert x_+\vert\right)}{\vert x_+\vert^{n+1}}\right],\label{eq:Omega_pos_con} \\
\Omega^{-(2n+1)}(x,x^\prime)&=\frac{\left(2\mu\right)^{-n}}{\sqrt{\pi}\Gamma\left(n+\frac{1}{2}\right)}\sum_{p=-\infty}^\infty\bigg[\vert x_-\vert^{n}K_{n}\left(\mu\vert x_-\vert\right)-\vert x_+\vert^{n}K_{n}\left(\mu\vert x_+\vert\right)\bigg]\label{eq:Omega_neg_con},
\end{align}
where $x_\pm=x \pm x^\prime-2pL$.
\subsection{Integral Representation and Resummation}
Since equations \eqref{eq:Omega_pos_con} and \eqref{eq:Omega_neg_con} contain infinite sums, they are not very handy for calculations. We use the integral representations of the Modified Bessel functions of the 2nd kind 
\begin{align}
K_\nu(z)&=\int_0^\infty dt\, e^{-z\cosh t}\cosh\left(\nu t\right),\\
K_\nu(z)&=\frac{\sqrt{\pi}\left(\frac{z}{2}\right)^\nu}{\Gamma\left(\nu+\frac{1}{2}\right)}\int_0^\infty dt\, e^{-z\cosh t}\sinh^{2\nu} t,\qquad
\end{align}
where $\re(z)>0$ and $\re(\nu)>-\frac{1}{2}$, and perform the summation over $p$ to obtain
\begin{align}
\hspace{-0.25cm}\Omega^{2n+1}(x,x^\prime)&=\frac{2\left(-\mu^2\right)^{n+1}}{\pi}\hspace{-0.15cm}\int_0^\infty \hspace{-0.1cm}dt\,\tilde{\Omega}^{2n+1}(x,x^\prime;t),\\
\hspace{-0.25cm}\Omega^{-1}(x,x^\prime)&=\frac{2}{\pi}\int_0^\infty \hspace{-0.1cm}dt\,\tilde{\Omega}^{-1}(x,x^\prime;t),
\end{align}
where
\begin{align}
\tilde{\Omega}_t^{2n+1}(x,x^\prime;t)&=\frac{\sinh\left[\mu(L-x_{>})\cosh t\right]\sinh\left(\mu x_{<}\cosh t\right)}{\sinh\left(\mu L\cosh t\right)\sinh^{-2(n+1)}t},\\
\tilde{\Omega}^{-1}_t(x,x^\prime;t)&=\frac{\sinh\left[\mu(L-x_{>})\cosh t\right]\sinh\left(\mu x_{<}\cosh t\right)}{\sinh\left(\mu L\cosh t\right)}.\label{eq:Omega_m1_t}
\end{align}
and $x_{>}:=\max(x,x^\prime)$, $x_{<}:=\min(x,x^\prime)$.
Notice that these integrals are convergent. A formula for $\Omega^{-(2n+1)}(x,x^\prime)$ for a generic value of $n$ is more complicated and we omit it, since it is not necessary for our analysis.

We calculate the kernel $\Omega^{-1}(x,x^\prime)$ using contour integration. First we extend the range of integration from $[0,\infty)$ to $(-\infty,\infty)$ using the fact that the integrand is an even function of $t$. Then we shift the integration variable $t\rightarrow t +i\pi$ to half of the integral to yield
\begin{equation}\label{eq:complex_int}
\Omega^{-1}(x,x^\prime)=\frac{1}{2\pi}\int_{-\infty}^\infty dt\,\tilde{\Omega}^{-1}(x,x^\prime;t)+\frac{1}{2\pi}\int_{\infty+i\pi}^{-\infty+i\pi}dt\,\tilde{\Omega}^{-1}(x,x^\prime;t),
\end{equation}
where $\tilde{\Omega}^{-1}(x,x^\prime;t)$ is given by \eqref{eq:Omega_m1_t}. Consider the contour integral in the complex plane along the rectangle defined by the edges $(-R,R)$, $(R,R+i\pi)$, $(R+i\pi,-R+i\pi)$ and $(-R+i\pi,R)$. It is easy to see that in the $R\rightarrow\infty$ limit, the sides that are parallel to the imaginary axis do not contribute, implying that it reduces to equation \eqref{eq:complex_int}. Its value is determined solely by its residues that lie at $z=\ln\left(\frac{k\pi}{\mu L}+\sqrt{1+\frac{k^2\pi^2}{\mu^2L^2}}\,\right)+i\frac{\pi}{2}$, where $k\in\mathbb{Z}^*$. Thus, we obtain
\begin{equation}
\Omega^{-1}(x,x^\prime)=\frac{2}{L}\sum_{k=1}^\infty\frac{1}{\omega_k}\sin\frac{k\pi x}{L}\sin\frac{k\pi x^\prime}{L},
\end{equation}
where, as expected, the eigenfrequencies $\omega_k$ are given by the equation
\begin{equation}
\omega_k=\left(\mu^2+\frac{k^2\pi^2}{L^2}\right)^{1/2}.
\end{equation}
Similarly, $\Omega^{2n+1}(x,x^\prime)$ is given by
\begin{equation}
\Omega^{2n+1}(x,x^\prime)=\frac{2}{L}\sum_{k=1}^\infty\omega_k^{2n+1}\sin\frac{k\pi x}{L}\sin\frac{k\pi x^\prime}{L}.
\end{equation}
It is easy to verify than in the massless case one obtains\footnote{Clausen functions of order $n$ are defined in terms of polylogarithms as
\begin{equation}
Cl_{n}(\theta) =\begin{cases}
\frac{i}{2}\left[L_{n}\left(e^{-i\theta}\right)-L_{n}\left(e^{i\theta}\right)\right],\quad n\quad\textrm{even},\\
\frac{1}{2}\left[L_{n}\left(e^{-i\theta}\right)+L_{n}\left(e^{i\theta}\right)\right],\quad n\quad\textrm{odd}.
\end{cases}
\end{equation}}
\begin{equation}
\begin{split}
\Omega^{2n+1}(x,x^\prime)&=\frac{2}{L}\sum_{k=1}^\infty\left[\frac{k\pi}{L}\right]^{2n+1}\sin\frac{k\pi x}{L}\sin\frac{k\pi y}{L}\\
&=\frac{\pi^{2n+1}}{L^{2n+2}}\left[Cl_{-(2n+1)}\left(\frac{\pi(x-y)}{L}\right)-Cl_{-(2n+1)}\left(\frac{\pi(x+y)}{L}\right)\right].
\end{split}
\end{equation}
The result can be analytically continued to $\mathbb{C}$ excluding even powers of $\Omega$, thus
\begin{equation}
\Omega^{-(2n+1)}(x,x^\prime)=\frac{L^{2n}}{\pi^{2n+1}}\left[Cl_{2n+1}\left(\frac{\pi(x-x^\prime)}{L}\right)-Cl_{2n+1}\left(\frac{\pi(x+x^\prime)}{L}\right)\right].
\end{equation}

As a consistency check, taking the massless limit of \eqref{eq:Omega_pos_con} and \eqref{eq:Omega_neg_con} one is able to verify that
\begin{align}
&\Omega(x,x^\prime)=\frac{\pi}{4L^2}\left(\frac{1}{\sin^2\frac{\pi(x+x^\prime)}{2L}}-\frac{1}{\sin^2\frac{\pi(x-x^\prime)}{2L}}\right),\\
&\Omega^{-1}(x,x^\prime)=\frac{1}{2\pi}\ln\frac{\sin^2\frac{\pi(x+x^\prime)}{2L}}{\sin^2\frac{\pi(x-x^\prime)}{2L}},
\end{align}
which coincide with \eqref{eq:kern_Omega} and \eqref{eq:Omega_inv}. Finally, the kernels are normalized since
\begin{equation}
\int_0^L dy\,\Omega^{-1}(x,y)\Omega(y,x^\prime)=\frac{2}{L}\sum_{n=1}^\infty\sin\frac{n\pi x}{L}\sin\frac{n\pi x^\prime}{L}=\delta(x-x^\prime).
\end{equation}

\subsection{Massless Limit Verification}
In this section we calculate the kernels $\Omega(x,x^\prime)$ and $\Omega^{-1}(x,x^\prime)$ in the massless limit directly from \eqref{eq:Omega_pos_con} and  \eqref{eq:Omega_neg_con}, without the contour integration of the previous section. For this purpose we need the expansions 
\begin{align}
K_0(\mu z)&=-\left(\gamma+\ln\frac{\mu z}{2}\right)+\mathcal{O}\left(\mu^2\right),\\
\mu^{n+1}K_{n+1}(\mu z)&=\frac{n!2^n}{z^{n+1}}+\mathcal{O}\left(\mu^2\right),\quad n\in\mathbb{N}.
\end{align}
Using them, it is straightforward to obtain
\begin{align}
\Omega^{2n+1}(x,x^\prime)&=\frac{(-1)^{n+1}(2n+1)!}{\pi}\sum_{p=-\infty}^\infty\left(x_-^{-2(n+1)}-x_+^{-2(n+1)}\right),
\\
\Omega^{-1}(x,x^\prime)&=-\frac{1}{\pi}\sum_{p=-\infty}^\infty\ln\left\vert \frac{x_-}{x_+}\right\vert,
\end{align}
where $x_\pm=x \pm x^\prime-2pL$. In particular the kernels $\Omega(x,x^\prime)$ and $\Omega^{-1}(x,x^\prime)$ read
\begin{align}
\Omega(x,x^\prime)&=-\frac{1}{4\pi L^2}\sum_{p=-\infty}^\infty\left[\left(\frac{x-x^\prime}{2L}-p\right)^{-2}-\left(\frac{x+x^\prime}{2L}-p\right)^{-2}\right],\\
\Omega^{-1}(x,x^\prime)&=-\frac{1}{\pi}\sum_{p=-\infty}^\infty\ln\left\vert \frac{\frac{x-x^\prime}{2L}-p}{\frac{x+x^\prime}{2L}-p}\right\vert.
\end{align}
Using the following infinite product
\begin{equation}\label{eq:infinite_prod}
\prod_{p=-\infty}^\infty\frac{\frac{x-x^\prime}{2L}-p}{\frac{x+x^\prime}{2L}-p}=\frac{x-x^\prime}{x+x^\prime}\prod_{p=1}^\infty\frac{\left(\frac{x-x^\prime}{2L}\right)^2-p^2}{\left(\frac{x+x^\prime}{2L}\right)^2-p^2}=\frac{\sin\frac{\pi\left(x-x^\prime\right)}{2L}}{\sin\frac{\pi\left(x+x^\prime\right)}{2L}}
\end{equation}
and sum
\begin{equation}
\sum_{p=-\infty}^\infty\frac{1}{\left(x-p\right)^2}=\frac{\pi^2}{\sin^2\left(\pi x\right)},
\end{equation}
we conclude that
\begin{align}
&\Omega(x,x^\prime)=\frac{\pi}{4L^2}\left[\frac{1}{\sin^2\frac{\pi (x+x^\prime)}{2L}}-\frac{1}{\sin^2\frac{\pi (x-x^\prime)}{2L}}\right],\\
&\Omega^{-1}(x,x^\prime)=-\frac{1}{\pi}\ln\left\vert\frac{\sin \frac{\pi\left(x-x^\prime\right)}{2L}}{\sin\frac{\pi\left(x+x^\prime\right)}{2L}}\right\vert,
\end{align}
which coincide with \eqref{eq:kern_Omega} and \eqref{eq:Omega_inv}.
\section{The Eigenvalue Problem and the Modular Hamiltonian\protect\label{app:appendix}}

In this appendix, we present the technical details of the derivation of the eigenvalues and eigenfunctions of the kernel $\tilde{M} \left( x , x^\prime \right)$ in the case of one-dimensional field theory defined on a finite interval and a single-point boundary separating the two subsystems, given by equation \eqref{eq:M_tilde_def}. Furthermore we present some technical details on the derivation of the modular Hamiltonian in the same configuration.

\subsection{The Eigenvalue Problem}
The eigenvalue problem that we would like to solve is given in equation \eqref{eq:eig_problem}. We introduce new coordinates $u=u(x)$ and $u^\prime=u(x^\prime)$ that facilitate its solution. In these coordinates the kernel becomes a sum of two terms: one that is a function of $u+u^\prime$ and another one that is the same function of $u-u^\prime$. This is achieved by defining the logarithms in equation \eqref{eq:M_tilde_def} as the new coordinates, namely\footnote{Had we considered the subsystem defined in $[0,\ell]$ we should substitute $\left(x-\ell\right)$ with $\left(\ell-x\right)$.}
\begin{align}
u&=u(x)=\ln\frac{\sin\frac{\pi\left(x+\ell\right)}{2L}}{\sin\frac{\pi\left(x-\ell\right)}{2L}}=\pi\, \Omega^{-1}(x,\ell),\\u^\prime&=u(x^\prime)=\ln\frac{\sin\frac{\pi\left(x^\prime+\ell\right)}{2L}}{\sin\frac{\pi\left(x^\prime-\ell\right)}{2L}}=\pi\,\Omega^{-1}(x^\prime,\ell).
\end{align}
These relations can be inverted as\footnote{For the subsystem defined in $[0,\ell]$ one should substitute $\coth$ with $\tanh$.}
\begin{align}
x &=\frac{2L}{\pi}\arctan\left(\tan\frac{\pi\ell}{2L}\coth\frac{u}{2}\right),\\ x^\prime &=\frac{2L}{\pi}\arctan\left(\tan\frac{\pi\ell}{2L}\coth\frac{u^\prime}{2}\right).
\end{align} 
Notice that
\begin{equation}
\lim_{u\rightarrow 0^+} x =L,\qquad \lim_{u\rightarrow +\infty} x = \ell.
\end{equation}
The above imply that in terms of the coordinates $u$ and $u^\prime$ the eigenvalue problem assumes the form
\begin{equation}\label{eq:eigen_prob_u}
\int_0^\infty \frac{du^\prime}{2\pi^2}\hspace{-0.04cm}\left[\frac{u^\prime-u}{\tanh\hspace{-0.02cm}\frac{u^\prime-u}{2}}-\frac{u^\prime+u}{\tanh\hspace{-0.02cm}\frac{u^\prime+u}{2}}\right]\hspace{-0.04cm}f(u^\prime)=\lambda f(u).
\end{equation}
The solution of this problem is 
\begin{equation}\label{eq:eigen_prob_u_sol}
f(u;\omega)=\sin(\omega u),\quad \lambda\left(\omega\right)=-\frac{1}{\sinh^2\left(\pi\omega\right)},
\end{equation}
where $\omega>0$. To prove this statement let us search for eigenfunctions of the form $f(u;\omega)$. We denote
\begin{equation}
I\equiv\int_0^\infty \frac{du^\prime}{2\pi^2}\left[\frac{u-u^\prime}{\tanh\frac{u-u^\prime}{2}}-\frac{u^\prime+u}{\tanh\frac{u+u^\prime}{2}}\right]\sin\left(\omega u^\prime\right).
\end{equation}
Then, we may split the two terms of the integrand and set $u^\prime\rightarrow-u^\prime$ on the first term. The two integrands coincide, thus we obtain a single integral, but the range of integration has been extended from $[0,\infty)$ to  $(-\infty,\infty)$. This in turn implies that we are free to shift the variable of integration, i.e.
\begin{equation}
I=-\int_{-\infty}^\infty \frac{du^\prime}{2\pi^2} \frac{u^\prime+u}{\tanh\frac{u+u^\prime}{2}}\sin\left(\omega u^\prime\right)=-\int_{-\infty}^\infty \frac{du^\prime}{2\pi^2} \frac{u^\prime}{\tanh \frac{u^\prime}{2}}\sin\left(\omega \left(u^\prime-u\right)\right).
\end{equation}
Finally, since the integral is defined in a symmetric domain, only the symmetric part of $\sin\left(\omega \left(u^\prime-u\right)\right)$ contributes, thus
\begin{equation}
I=\sin\left(\omega u\right)\int_{-\infty}^\infty \frac{du^\prime}{2\pi^2} \frac{u^\prime}{\tanh \frac{u^\prime}{2}}\cos\left(\omega u^\prime\right).
\end{equation}
This integral can be calculated using contour integration and one ends up with
\begin{equation}
I=-\frac{1}{\sinh^2\left(\pi\omega\right)}\sin\left(\omega u\right).
\end{equation}

The kernel $\tilde{M}$ is not symmetric, which implies that its left and right eigenfunctions do not coincide. As a result, in order to obtain the spectral decomposition of $\tilde{M}$ we have to specify the left eigenfunctions too. It turns out that this is straightforward. The eigenvalue problem for the left eigenfunctions $g(u)$ reads
\begin{equation}
\int_0^\infty \frac{du}{2\pi^2}\frac{\cosh u^\prime-\cos\frac{\pi \ell}{L}}{\cosh u-\cos\frac{\pi \ell}{L}}\left(\frac{u^\prime-u}{\tanh\frac{u^\prime-u}{2}}-\frac{u^\prime+u}{\tanh\frac{u^\prime+u}{2}}\right)g(u)=-\frac{1}{\sinh^2\left(\pi\omega\right)} g(u^\prime).
\end{equation}
Thus, trivially, the eigenfunctions read
\begin{equation}\label{eq:g_of_f}
g(u)=\left(\cosh u-\cos\frac{\pi \ell}{L}\right)f(u).
\end{equation}
Notice that
\begin{equation}\label{eq:dx_du}
\left\vert\frac{dx}{du}\right\vert=\frac{L}{\pi}\frac{\sin\frac{\pi\ell}{L}}{\cosh u- \cos\frac{\pi\ell}{L}},
\end{equation}
so this extra factor in equation \eqref{eq:g_of_f} is related to the change of variable from $x$ to $u$.

We have obtained the left and right eigenfunctions separately. We need to fix their normalization. Since we have parametrized the eigenfunctions using the continuous parameter $\omega$, we specify the normalization factor $h(\omega)$ demanding that
\begin{equation}
h(\omega)\int_{\ell}^{L}dx\, g(u(x);\omega)f(u(x);\omega^\prime)=\delta\left(\omega-\omega^\prime\right).
\end{equation}
Changing the variable of integration to $u$, we obtain
\begin{equation}
\frac{L}{\pi} h\left(\omega\right)\sin\frac{\pi \ell}{L}\int_0^\infty du \sin(\omega u)\sin(\omega^\prime u)=\delta\left(\omega-\omega^\prime\right)=\frac{2}{\pi}\int_0^\infty du \sin(\omega u)\sin(\omega^\prime u).
\end{equation}
As a result, the normalization factor is
\begin{equation}
h\left(\omega\right)=\frac{2}{L}\frac{1}{\sin\frac{\pi \ell}{L}}.
\end{equation}
Putting everything together, the spectral decomposition of the kernel $\tilde{M}$ is given by \eqref{eq:Mtilde_decomp}.

\subsection{The Modular Hamiltonian}

In this appendix we derive the closed form of the kernels $\tilde{K}(x,y)$ and $\tilde{V}(x,y)$, which are given by \eqref{eq:tilde_K_init} and \eqref{eq:tilde_V_init}. This calculation results in the modular Hamiltonian \eqref{eq:Modular_Large}. 

Recall that we use the notation $u_x=u(x)$, $u_y=u(y)$ and $u^\prime=u(x^\prime)$. First we perform the integration over $u^\prime$. We express the logarithm in \eqref{eq:tilde_K_init} as difference of two logarithms and split the integral into two terms. Then we perform the change of variables $u^\prime\rightarrow -u^\prime$ on the second integral only. After this change of variable the integrands of the two integrals become identical. We combine both integrals into a single integral that has range of integration from $-\infty$ to $+\infty$. Then, we perform the change of variable $u^\prime\rightarrow u^\prime + u_y$ to yield
\begin{equation}
\tilde{K}(x,y)=-\frac{2}{\pi}\int_0^\infty d\omega\, \frac{\omega\sin(\omega u_x)}{\coth\left(\pi\omega\right)}\int_{-\infty}^\infty du^\prime\sin\left(\omega \left(u^\prime+u_y\right)\right)\ln\left\vert\sinh\frac{u^\prime}{2}\right\vert.
\end{equation}
Only the symmetric part of $\sin\left(\omega \left(u^\prime+u_y\right)\right)$ contributes to the result, thus we obtain
\begin{equation}
\tilde{K}(x,y)=-\frac{4}{\pi}\int_0^\infty d\omega\, \frac{\omega\sin(\omega u_x)\sin\left(\omega u_y\right)}{\coth\left(\pi\omega\right)}\int_{0}^{\infty} du^\prime\cos\left(\omega u^\prime\right)\ln\sinh\frac{u^\prime}{2}.
\end{equation}
The $u^\prime$ integral can be calculated as\footnote{The following Fourier transforms are required:
\begin{align}
&\int_{0}^{\infty} du^\prime\cos\left(\omega u^\prime\right)\log\left[1-\exp\left(-u^\prime\right)\right]=\frac{1-\pi \omega\coth\left(\pi\omega\right)}{2\omega^2},\\
&\int_{0}^{\infty} du^\prime\cos\left(\omega u^\prime\right)u^\prime=\frac{d}{d\omega}\int_{0}^{\infty} du^\prime\sin\left(\omega u^\prime\right)=-\frac{1}{\omega^2},\\
&\int_{0}^{\infty} du^\prime\cos\left(\omega u^\prime\right)=\pi\delta(\omega).
\end{align}}
\begin{equation}\label{eq:log_ft}
-\frac{1}{\pi}\int_{0}^{\infty} du^\prime\cos\left(\omega u^\prime\right)\ln\sinh\frac{u^\prime}{2}=\frac{1}{2\omega}\coth\left(\pi\omega\right)+ \delta(\omega)\ln2.
\end{equation}
Thus, putting everything together, the kernel reads
\begin{equation}
\tilde{K}(x,y)=2\int_0^\infty d\omega\sin(\omega u_x)\sin\left(\omega u_y\right)=\pi\delta\left(u_x-u_y\right).
\end{equation}
Restoring the original coordinates, it turns out that the kernel $\tilde{K}(x,y)$ is given by \eqref{eq:kernel_K_final}.

We treat in a similar manner the kernel $\tilde{V}(x,y)$, which is given by \eqref{eq:tilde_V_init}. We perform the integral over $u^\prime$ first. We implement the same strategy in order to express this integral as an integral of a single term over the region $(-\infty,\infty)$ and then we shift the coordinate $u^\prime$. Doing so, the integral assumes the form
\begin{equation}
\int_0^\infty du^\prime \sin(\omega u^\prime) \left(\frac{1}{\sinh^2\frac{u_x+u^\prime}{2}}-\frac{1}{\sinh^2\frac{u_x-u^\prime}{2}}\right)=\int_{-\infty}^\infty du^\prime  \frac{\sin(\omega \left(u^\prime-u_x\right))}{\sinh^2\frac{u^\prime}{2}}.
\end{equation}
Only the symmetric part of $\sin(\omega \left(u^\prime-u_x\right))$ contributes. We obtain
\begin{equation}
\int_0^\infty du^\prime \sin(\omega u^\prime) \left(\frac{1}{\sinh^2\frac{u_x+u^\prime}{2}}-\frac{1}{\sinh^2\frac{u_x-u^\prime}{2}}\right)=-2\sin\left(\omega u_x\right)\int_{0}^\infty du^\prime \frac{\cos\left(\omega u^\prime\right)}{\sinh^2\frac{u^\prime}{2}}.
\end{equation}
This integral can be calculated in the complex plane using contour integration\footnote{We express the cosine as a sum of two exponentials and then set $u^\prime\rightarrow-u^\prime$ in the term $\exp(-i \omega u^\prime)$. We obtain an integral of a single exponential from $-\infty$ to $+\infty$. We can close the contour using the upper half plane. The integral has poles on the imaginary axis at $u^\prime= 2 i n\pi$. Using the contour that does not include the pole at $u^\prime = 0$, we obtain
\begin{equation}
\int_{-\infty}^{\infty}du^\prime\frac{e^{i\omega u^\prime}}{\sinh^2\frac{u^\prime}{2}}=\frac{1}{2}\left(-8\pi\omega\right)+\sum_{n=1}^\infty \left(-8\pi \omega e^{-2n \pi \omega}\right)=-4\pi\omega\coth\left(\pi\omega\right),
\end{equation}
where the first term is the contribution of the integration around the pole at $u^\prime=0$ after dropping the divergent term.}, yielding
\begin{equation}
\int_0^\infty du^\prime \sin(\omega u^\prime) \left(\frac{1}{\sinh^2\frac{u_x+u^\prime}{2}}-\frac{1}{\sinh^2\frac{u_x-u^\prime}{2}}\right)=4\pi\omega\sin\left(\omega u_x\right)\coth\left(\pi\omega\right).
\end{equation}
Substituting in \eqref{eq:tilde_V_init} and using \eqref{eq:dx_du} the kernel $\tilde{V}(x,y)$ reads
\begin{equation}
\tilde{V}(x,y)=2\frac{d u_x}{dx}\frac{d u_y}{dy}\int_0^\infty d\omega \omega^2\sin(\omega u_y)\sin\left(\omega u_x\right)=-\frac{d u_x}{dx}\frac{d u_y}{dy}\frac{d^2}{d u_x^2}\delta\left(u_x-u_y\right).
\end{equation}
Using properties of the delta function, it is straightforward to show that $\tilde{V}(x,y)$ assumes the form \eqref{eq:kernel_V_final}.
\bibliographystyle{JHEP}
\bibliography{review_final}

\providecommand{\href}[2]{#2}\begingroup\raggedright\begin{thebibliography}{100}

\bibitem{Einstein:1935rr}
A.~Einstein, B.~Podolsky and N.~Rosen, \emph{{Can quantum mechanical
  description of physical reality be considered complete?}},
  \href{https://doi.org/10.1103/PhysRev.47.777}{\emph{Phys. Rev.} {\bfseries
  47} (1935) 777}.

\bibitem{schrodinger_1935}
E.~Schrödinger, \emph{Discussion of probability relations between separated
  systems}, \href{https://doi.org/10.1017/S0305004100013554}{\emph{Mathematical
  Proceedings of the Cambridge Philosophical Society} {\bfseries 31} (1935)
  555–563}.

\bibitem{Chruscinski:2014oca}
D.~Chruscinski and G.~Sarbicki, \emph{{Entanglement witnesses: construction,
  analysis and classification}},
  \href{https://doi.org/10.1088/1751-8113/47/48/483001}{\emph{J. Phys. A}
  {\bfseries 47} (2014) 483001}
  [\href{https://arxiv.org/abs/1402.2413}{{\ttfamily 1402.2413}}].

\bibitem{Vidal:2002rm}
G.~Vidal, J.I.~Latorre, E.~Rico and A.~Kitaev, \emph{{Entanglement in quantum
  critical phenomena}},
  \href{https://doi.org/10.1103/PhysRevLett.90.227902}{\emph{Phys. Rev. Lett.}
  {\bfseries 90} (2003) 227902}
  [\href{https://arxiv.org/abs/quant-ph/0211074}{{\ttfamily
  quant-ph/0211074}}].

\bibitem{Levin:2006zz}
M.~Levin and X.-G.~Wen, \emph{{Detecting Topological Order in a Ground State
  Wave Function}},
  \href{https://doi.org/10.1103/PhysRevLett.96.110405}{\emph{Phys. Rev. Lett.}
  {\bfseries 96} (2006) 110405}
  [\href{https://arxiv.org/abs/cond-mat/0510613}{{\ttfamily
  cond-mat/0510613}}].

\bibitem{Kitaev:2005dm}
A.~Kitaev and J.~Preskill, \emph{{Topological entanglement entropy}},
  \href{https://doi.org/10.1103/PhysRevLett.96.110404}{\emph{Phys. Rev. Lett.}
  {\bfseries 96} (2006) 110404}
  [\href{https://arxiv.org/abs/hep-th/0510092}{{\ttfamily hep-th/0510092}}].

\bibitem{Sorkin:1984kjy}
R.D.~Sorkin, \emph{{1983 paper on entanglement entropy: ''On the Entropy of the
  Vacuum outside a Horizon''}},  in \emph{{10th International Conference on
  General Relativity and Gravitation}}, vol.~2, pp.~734--736, 1984
  [\href{https://arxiv.org/abs/1402.3589}{{\ttfamily 1402.3589}}].

\bibitem{Bombelli:1986rw}
L.~Bombelli, R.K.~Koul, J.~Lee and R.D.~Sorkin, \emph{{A Quantum Source of
  Entropy for Black Holes}},
  \href{https://doi.org/10.1103/PhysRevD.34.373}{\emph{Phys. Rev. D} {\bfseries
  34} (1986) 373}.

\bibitem{Srednicki:1993im}
M.~Srednicki, \emph{{Entropy and area}},
  \href{https://doi.org/10.1103/PhysRevLett.71.666}{\emph{Phys. Rev. Lett.}
  {\bfseries 71} (1993) 666}
  [\href{https://arxiv.org/abs/hep-th/9303048}{{\ttfamily hep-th/9303048}}].

\bibitem{Bekenstein:1973ur}
J.D.~Bekenstein, \emph{{Black holes and entropy}},
  \href{https://doi.org/10.1103/PhysRevD.7.2333}{\emph{Phys. Rev. D} {\bfseries
  7} (1973) 2333}.

\bibitem{Bardeen:1973gs}
J.M.~Bardeen, B.~Carter and S.W.~Hawking, \emph{{The Four laws of black hole
  mechanics}}, \href{https://doi.org/10.1007/BF01645742}{\emph{Commun. Math.
  Phys.} {\bfseries 31} (1973) 161}.

\bibitem{Callan:1994py}
C.G.~Callan, Jr. and F.~Wilczek, \emph{{On geometric entropy}},
  \href{https://doi.org/10.1016/0370-2693(94)91007-3}{\emph{Phys. Lett. B}
  {\bfseries 333} (1994) 55}
  [\href{https://arxiv.org/abs/hep-th/9401072}{{\ttfamily hep-th/9401072}}].

\bibitem{Holzhey:1994we}
C.~Holzhey, F.~Larsen and F.~Wilczek, \emph{{Geometric and renormalized entropy
  in conformal field theory}},
  \href{https://doi.org/10.1016/0550-3213(94)90402-2}{\emph{Nucl. Phys. B}
  {\bfseries 424} (1994) 443}
  [\href{https://arxiv.org/abs/hep-th/9403108}{{\ttfamily hep-th/9403108}}].

\bibitem{Larsen:1994yt}
F.~Larsen and F.~Wilczek, \emph{{Geometric entropy, wave functionals, and
  fermions}}, \href{https://doi.org/10.1006/aphy.1995.1100}{\emph{Annals Phys.}
  {\bfseries 243} (1995) 280}
  [\href{https://arxiv.org/abs/hep-th/9408089}{{\ttfamily hep-th/9408089}}].

\bibitem{renyi1961}
A.~R{\'e}nyi, \emph{On measures of entropy and information},  in
  \emph{Proceedings of the Fourth Berkeley Symposium on Mathematical Statistics
  and Probability, Volume 1: Contributions to the Theory of Statistics},
  vol.~4, pp.~547--562, University of California Press, 1961.

\bibitem{Calabrese:2004eu}
P.~Calabrese and J.L.~Cardy, \emph{{Entanglement entropy and quantum field
  theory}}, \href{https://doi.org/10.1088/1742-5468/2004/06/P06002}{\emph{J.
  Stat. Mech.} {\bfseries 0406} (2004) P06002}
  [\href{https://arxiv.org/abs/hep-th/0405152}{{\ttfamily hep-th/0405152}}].

\bibitem{Casini:2004bw}
H.~Casini and M.~Huerta, \emph{{A Finite entanglement entropy and the
  c-theorem}},
  \href{https://doi.org/10.1016/j.physletb.2004.08.072}{\emph{Phys. Lett. B}
  {\bfseries 600} (2004) 142}
  [\href{https://arxiv.org/abs/hep-th/0405111}{{\ttfamily hep-th/0405111}}].

\bibitem{Balasubramanian:2011wt}
V.~Balasubramanian, M.B.~McDermott and M.~Van~Raamsdonk, \emph{{Momentum-space
  entanglement and renormalization in quantum field theory}},
  \href{https://doi.org/10.1103/PhysRevD.86.045014}{\emph{Phys. Rev. D}
  {\bfseries 86} (2012) 045014}
  [\href{https://arxiv.org/abs/1108.3568}{{\ttfamily 1108.3568}}].

\bibitem{Calabrese:2009qy}
P.~Calabrese and J.~Cardy, \emph{{Entanglement entropy and conformal field
  theory}}, \href{https://doi.org/10.1088/1751-8113/42/50/504005}{\emph{J.
  Phys. A} {\bfseries 42} (2009) 504005}
  [\href{https://arxiv.org/abs/0905.4013}{{\ttfamily 0905.4013}}].

\bibitem{Casini:2009sr}
H.~Casini and M.~Huerta, \emph{{Entanglement entropy in free quantum field
  theory}}, \href{https://doi.org/10.1088/1751-8113/42/50/504007}{\emph{J.
  Phys. A} {\bfseries 42} (2009) 504007}
  [\href{https://arxiv.org/abs/0905.2562}{{\ttfamily 0905.2562}}].

\bibitem{Casini:2022rlv}
H.~Casini and M.~Huerta, \emph{{Lectures on entanglement in quantum field
  theory}}, \href{https://doi.org/10.22323/1.403.0002}{\emph{PoS} {\bfseries
  TASI2021} (2023) 002} [\href{https://arxiv.org/abs/2201.13310}{{\ttfamily
  2201.13310}}].

\bibitem{Amico:2007ag}
L.~Amico, R.~Fazio, A.~Osterloh and V.~Vedral, \emph{{Entanglement in many-body
  systems}}, \href{https://doi.org/10.1103/RevModPhys.80.517}{\emph{Rev. Mod.
  Phys.} {\bfseries 80} (2008) 517}
  [\href{https://arxiv.org/abs/quant-ph/0703044}{{\ttfamily
  quant-ph/0703044}}].

\bibitem{Eisert:2008ur}
J.~Eisert, M.~Cramer and M.B.~Plenio, \emph{{Area laws for the entanglement
  entropy - a review}},
  \href{https://doi.org/10.1103/RevModPhys.82.277}{\emph{Rev. Mod. Phys.}
  {\bfseries 82} (2010) 277} [\href{https://arxiv.org/abs/0808.3773}{{\ttfamily
  0808.3773}}].

\bibitem{Maldacena:1997re}
J.M.~Maldacena, \emph{{The Large N limit of superconformal field theories and
  supergravity}}, \href{https://doi.org/10.4310/ATMP.1998.v2.n2.a1}{\emph{Adv.
  Theor. Math. Phys.} {\bfseries 2} (1998) 231}
  [\href{https://arxiv.org/abs/hep-th/9711200}{{\ttfamily hep-th/9711200}}].

\bibitem{Gubser:1998bc}
S.S.~Gubser, I.R.~Klebanov and A.M.~Polyakov, \emph{{Gauge theory correlators
  from noncritical string theory}},
  \href{https://doi.org/10.1016/S0370-2693(98)00377-3}{\emph{Phys. Lett. B}
  {\bfseries 428} (1998) 105}
  [\href{https://arxiv.org/abs/hep-th/9802109}{{\ttfamily hep-th/9802109}}].

\bibitem{Witten:1998qj}
E.~Witten, \emph{{Anti-de Sitter space and holography}},
  \href{https://doi.org/10.4310/ATMP.1998.v2.n2.a2}{\emph{Adv. Theor. Math.
  Phys.} {\bfseries 2} (1998) 253}
  [\href{https://arxiv.org/abs/hep-th/9802150}{{\ttfamily hep-th/9802150}}].

\bibitem{Ryu:2006bv}
S.~Ryu and T.~Takayanagi, \emph{{Holographic derivation of entanglement entropy
  from AdS/CFT}},
  \href{https://doi.org/10.1103/PhysRevLett.96.181602}{\emph{Phys. Rev. Lett.}
  {\bfseries 96} (2006) 181602}
  [\href{https://arxiv.org/abs/hep-th/0603001}{{\ttfamily hep-th/0603001}}].

\bibitem{Ryu:2006ef}
S.~Ryu and T.~Takayanagi, \emph{{Aspects of Holographic Entanglement Entropy}},
  \href{https://doi.org/10.1088/1126-6708/2006/08/045}{\emph{JHEP} {\bfseries
  08} (2006) 045} [\href{https://arxiv.org/abs/hep-th/0605073}{{\ttfamily
  hep-th/0605073}}].

\bibitem{Hubeny:2007xt}
V.E.~Hubeny, M.~Rangamani and T.~Takayanagi, \emph{{A Covariant holographic
  entanglement entropy proposal}},
  \href{https://doi.org/10.1088/1126-6708/2007/07/062}{\emph{JHEP} {\bfseries
  07} (2007) 062} [\href{https://arxiv.org/abs/0705.0016}{{\ttfamily
  0705.0016}}].

\bibitem{Lewkowycz:2013nqa}
A.~Lewkowycz and J.~Maldacena, \emph{{Generalized gravitational entropy}},
  \href{https://doi.org/10.1007/JHEP08(2013)090}{\emph{JHEP} {\bfseries 08}
  (2013) 090} [\href{https://arxiv.org/abs/1304.4926}{{\ttfamily 1304.4926}}].

\bibitem{Page:1993wv}
D.N.~Page, \emph{{Information in black hole radiation}},
  \href{https://doi.org/10.1103/PhysRevLett.71.3743}{\emph{Phys. Rev. Lett.}
  {\bfseries 71} (1993) 3743}
  [\href{https://arxiv.org/abs/hep-th/9306083}{{\ttfamily hep-th/9306083}}].

\bibitem{Penington:2019npb}
G.~Penington, \emph{{Entanglement Wedge Reconstruction and the Information
  Paradox}}, \href{https://doi.org/10.1007/JHEP09(2020)002}{\emph{JHEP}
  {\bfseries 09} (2020) 002}
  [\href{https://arxiv.org/abs/1905.08255}{{\ttfamily 1905.08255}}].

\bibitem{Almheiri:2019psf}
A.~Almheiri, N.~Engelhardt, D.~Marolf and H.~Maxfield, \emph{{The entropy of
  bulk quantum fields and the entanglement wedge of an evaporating black
  hole}}, \href{https://doi.org/10.1007/JHEP12(2019)063}{\emph{JHEP} {\bfseries
  12} (2019) 063} [\href{https://arxiv.org/abs/1905.08762}{{\ttfamily
  1905.08762}}].

\bibitem{Klebanov:2007ws}
I.R.~Klebanov, D.~Kutasov and A.~Murugan, \emph{{Entanglement as a probe of
  confinement}},
  \href{https://doi.org/10.1016/j.nuclphysb.2007.12.017}{\emph{Nucl. Phys. B}
  {\bfseries 796} (2008) 274}
  [\href{https://arxiv.org/abs/0709.2140}{{\ttfamily 0709.2140}}].

\bibitem{Nishioka:2009un}
T.~Nishioka, S.~Ryu and T.~Takayanagi, \emph{{Holographic Entanglement Entropy:
  An Overview}}, \href{https://doi.org/10.1088/1751-8113/42/50/504008}{\emph{J.
  Phys. A} {\bfseries 42} (2009) 504008}
  [\href{https://arxiv.org/abs/0905.0932}{{\ttfamily 0905.0932}}].

\bibitem{Takayanagi:2012kg}
T.~Takayanagi, \emph{{Entanglement Entropy from a Holographic Viewpoint}},
  \href{https://doi.org/10.1088/0264-9381/29/15/153001}{\emph{Class. Quant.
  Grav.} {\bfseries 29} (2012) 153001}
  [\href{https://arxiv.org/abs/1204.2450}{{\ttfamily 1204.2450}}].

\bibitem{VanRaamsdonk:2016exw}
M.~Van~Raamsdonk, \emph{{Lectures on Gravity and Entanglement}},  in
  \emph{{Theoretical Advanced Study Institute in Elementary Particle Physics}:
  {New Frontiers in Fields and Strings}}, pp.~297--351, 2017,
  \href{https://doi.org/10.1142/9789813149441_0005}{DOI}
  [\href{https://arxiv.org/abs/1609.00026}{{\ttfamily 1609.00026}}].

\bibitem{Rangamani:2016dms}
M.~Rangamani and T.~Takayanagi, \emph{{Holographic Entanglement Entropy}},
  vol.~931, Springer (2017),
  \href{https://doi.org/10.1007/978-3-319-52573-0}{10.1007/978-3-319-52573-0},
  [\href{https://arxiv.org/abs/1609.01287}{{\ttfamily 1609.01287}}].

\bibitem{Nishioka:2018khk}
T.~Nishioka, \emph{{Entanglement entropy: holography and renormalization
  group}}, \href{https://doi.org/10.1103/RevModPhys.90.035007}{\emph{Rev. Mod.
  Phys.} {\bfseries 90} (2018) 035007}
  [\href{https://arxiv.org/abs/1801.10352}{{\ttfamily 1801.10352}}].

\bibitem{Jacobson:1995ab}
T.~Jacobson, \emph{{Thermodynamics of space-time: The Einstein equation of
  state}}, \href{https://doi.org/10.1103/PhysRevLett.75.1260}{\emph{Phys. Rev.
  Lett.} {\bfseries 75} (1995) 1260}
  [\href{https://arxiv.org/abs/gr-qc/9504004}{{\ttfamily gr-qc/9504004}}].

\bibitem{Verlinde:2010hp}
E.P.~Verlinde, \emph{{On the Origin of Gravity and the Laws of Newton}},
  \href{https://doi.org/10.1007/JHEP04(2011)029}{\emph{JHEP} {\bfseries 04}
  (2011) 029} [\href{https://arxiv.org/abs/1001.0785}{{\ttfamily 1001.0785}}].

\bibitem{Jacobson:2012yt}
T.~Jacobson, \emph{{Gravitation and vacuum entanglement entropy}},
  \href{https://doi.org/10.1142/S0218271812420060}{\emph{Int. J. Mod. Phys. D}
  {\bfseries 21} (2012) 1242006}
  [\href{https://arxiv.org/abs/1204.6349}{{\ttfamily 1204.6349}}].

\bibitem{Jacobson:2015hqa}
T.~Jacobson, \emph{{Entanglement Equilibrium and the Einstein Equation}},
  \href{https://doi.org/10.1103/PhysRevLett.116.201101}{\emph{Phys. Rev. Lett.}
  {\bfseries 116} (2016) 201101}
  [\href{https://arxiv.org/abs/1505.04753}{{\ttfamily 1505.04753}}].

\bibitem{VanRaamsdonk:2010pw}
M.~Van~Raamsdonk, \emph{{Building up spacetime with quantum entanglement}},
  \href{https://doi.org/10.1142/S0218271810018529}{\emph{Gen. Rel. Grav.}
  {\bfseries 42} (2010) 2323}
  [\href{https://arxiv.org/abs/1005.3035}{{\ttfamily 1005.3035}}].

\bibitem{VanRaamsdonk:2009ar}
M.~Van~Raamsdonk, \emph{{Comments on quantum gravity and entanglement}},
  \href{https://arxiv.org/abs/0907.2939}{{\ttfamily 0907.2939}}.

\bibitem{Lashkari:2013koa}
N.~Lashkari, M.B.~McDermott and M.~Van~Raamsdonk, \emph{{Gravitational dynamics
  from entanglement 'thermodynamics'}},
  \href{https://doi.org/10.1007/JHEP04(2014)195}{\emph{JHEP} {\bfseries 04}
  (2014) 195} [\href{https://arxiv.org/abs/1308.3716}{{\ttfamily 1308.3716}}].

\bibitem{Faulkner:2013ica}
T.~Faulkner, M.~Guica, T.~Hartman, R.C.~Myers and M.~Van~Raamsdonk,
  \emph{{Gravitation from Entanglement in Holographic CFTs}},
  \href{https://doi.org/10.1007/JHEP03(2014)051}{\emph{JHEP} {\bfseries 03}
  (2014) 051} [\href{https://arxiv.org/abs/1312.7856}{{\ttfamily 1312.7856}}].

\bibitem{Maldacena:2013xja}
J.~Maldacena and L.~Susskind, \emph{{Cool horizons for entangled black holes}},
  \href{https://doi.org/10.1002/prop.201300020}{\emph{Fortsch. Phys.}
  {\bfseries 61} (2013) 781} [\href{https://arxiv.org/abs/1306.0533}{{\ttfamily
  1306.0533}}].

\bibitem{Susskind:2017ney}
L.~Susskind, \emph{{Dear Qubitzers, GR=QM}},
  \href{https://arxiv.org/abs/1708.03040}{{\ttfamily 1708.03040}}.

\bibitem{Blanco:2013joa}
D.D.~Blanco, H.~Casini, L.-Y.~Hung and R.C.~Myers, \emph{{Relative Entropy and
  Holography}}, \href{https://doi.org/10.1007/JHEP08(2013)060}{\emph{JHEP}
  {\bfseries 08} (2013) 060} [\href{https://arxiv.org/abs/1305.3182}{{\ttfamily
  1305.3182}}].

\bibitem{Page:1993df}
D.N.~Page, \emph{{Average entropy of a subsystem}},
  \href{https://doi.org/10.1103/PhysRevLett.71.1291}{\emph{Phys. Rev. Lett.}
  {\bfseries 71} (1993) 1291}
  [\href{https://arxiv.org/abs/gr-qc/9305007}{{\ttfamily gr-qc/9305007}}].

\bibitem{Moore:1991ks}
G.W.~Moore and N.~Read, \emph{{Nonabelions in the fractional quantum Hall
  effect}}, \href{https://doi.org/10.1016/0550-3213(91)90407-O}{\emph{Nucl.
  Phys. B} {\bfseries 360} (1991) 362}.

\bibitem{Li:2008kda}
H.~Li and F.~Haldane, \emph{{Entanglement Spectrum as a Generalization of
  Entanglement Entropy: Identification of Topological Order in Non-Abelian
  Fractional Quantum Hall Effect States}},
  \href{https://doi.org/10.1103/PhysRevLett.101.010504}{\emph{Phys. Rev. Lett.}
  {\bfseries 101} (2008) 010504}
  [\href{https://arxiv.org/abs/0805.0332}{{\ttfamily 0805.0332}}].

\bibitem{Bisognano:1975ih}
J.J.~Bisognano and E.H.~Wichmann, \emph{{On the Duality Condition for a
  Hermitian Scalar Field}}, \href{https://doi.org/10.1063/1.522605}{\emph{J.
  Math. Phys.} {\bfseries 16} (1975) 985}.

\bibitem{Bisognano:1976za}
J.J.~Bisognano and E.H.~Wichmann, \emph{{On the Duality Condition for Quantum
  Fields}}, \href{https://doi.org/10.1063/1.522898}{\emph{J. Math. Phys.}
  {\bfseries 17} (1976) 303}.

\bibitem{Casini:2011kv}
H.~Casini, M.~Huerta and R.C.~Myers, \emph{{Towards a derivation of holographic
  entanglement entropy}},
  \href{https://doi.org/10.1007/JHEP05(2011)036}{\emph{JHEP} {\bfseries 05}
  (2011) 036} [\href{https://arxiv.org/abs/1102.0440}{{\ttfamily 1102.0440}}].

\bibitem{Hislop:1981uh}
P.D.~Hislop and R.~Longo, \emph{{Modular Structure of the Local Algebras
  Associated With the Free Massless Scalar Field Theory}},
  \href{https://doi.org/10.1007/BF01208372}{\emph{Commun. Math. Phys.}
  {\bfseries 84} (1982) 71}.

\bibitem{Katsinis:2017qzh}
D.~Katsinis and G.~Pastras, \emph{{An Inverse Mass Expansion for Entanglement
  Entropy in Free Massive Scalar Field Theory}},
  \href{https://doi.org/10.1140/epjc/s10052-018-5596-4}{\emph{Eur. Phys. J. C}
  {\bfseries 78} (2018) 282}
  [\href{https://arxiv.org/abs/1711.02618}{{\ttfamily 1711.02618}}].

\bibitem{Katsinis:2019vhk}
D.~Katsinis and G.~Pastras, \emph{{Area Law Behaviour of Mutual Information at
  Finite Temperature}},  \href{https://arxiv.org/abs/1907.04817}{{\ttfamily
  1907.04817}}.

\bibitem{Katsinis:2019lis}
D.~Katsinis and G.~Pastras, \emph{{An Inverse Mass Expansion for the Mutual
  Information in Free Scalar QFT at Finite Temperature}},
  \href{https://doi.org/10.1007/JHEP02(2020)091}{\emph{JHEP} {\bfseries 02}
  (2020) 091} [\href{https://arxiv.org/abs/1907.08508}{{\ttfamily
  1907.08508}}].

\bibitem{Benedict:1995yp}
E.~Benedict and S.-Y.~Pi, \emph{{Entanglement entropy of nontrivial states}},
  \href{https://doi.org/10.1006/aphy.1996.0007}{\emph{Annals Phys.} {\bfseries
  245} (1996) 209} [\href{https://arxiv.org/abs/hep-th/9505121}{{\ttfamily
  hep-th/9505121}}].

\bibitem{Katsinis:2022fxu}
D.~Katsinis and G.~Pastras, \emph{{Entanglement in harmonic systems at coherent
  states}},  \href{https://arxiv.org/abs/2206.05781}{{\ttfamily 2206.05781}}.

\bibitem{Bianchi:2015fra}
E.~Bianchi, L.~Hackl and N.~Yokomizo, \emph{{Entanglement entropy of squeezed
  vacua on a lattice}},
  \href{https://doi.org/10.1103/PhysRevD.92.085045}{\emph{Phys. Rev. D}
  {\bfseries 92} (2015) 085045}
  [\href{https://arxiv.org/abs/1507.01567}{{\ttfamily 1507.01567}}].

\bibitem{Katsinis:2023hqn}
D.~Katsinis, G.~Pastras and N.~Tetradis, \emph{{Entanglement of Harmonic
  Systems in Squeezed States}},
  \href{https://arxiv.org/abs/2304.04241}{{\ttfamily 2304.04241}}.

\bibitem{Katsinis:2024sek}
D.~Katsinis, G.~Pastras and N.~Tetradis, \emph{{Entanglement Entropy of a
  Scalar Field in a Squeezed State}},
  \href{https://arxiv.org/abs/2403.03136}{{\ttfamily 2403.03136}}.

\bibitem{Cardy:2016fqc}
J.~Cardy and E.~Tonni, \emph{{Entanglement hamiltonians in two-dimensional
  conformal field theory}},
  \href{https://doi.org/10.1088/1742-5468/2016/12/123103}{\emph{J. Stat. Mech.}
  {\bfseries 1612} (2016) 123103}
  [\href{https://arxiv.org/abs/1608.01283}{{\ttfamily 1608.01283}}].

\bibitem{Arias:2016nip}
R.~Arias, D.~Blanco, H.~Casini and M.~Huerta, \emph{{Local temperatures and
  local terms in modular Hamiltonians}},
  \href{https://doi.org/10.1103/PhysRevD.95.065005}{\emph{Phys. Rev. D}
  {\bfseries 95} (2017) 065005}
  [\href{https://arxiv.org/abs/1611.08517}{{\ttfamily 1611.08517}}].

\bibitem{Arias:2017dda}
R.~Arias, H.~Casini, M.~Huerta and D.~Pontello, \emph{{Anisotropic Unruh
  temperatures}}, \href{https://doi.org/10.1103/PhysRevD.96.105019}{\emph{Phys.
  Rev. D} {\bfseries 96} (2017) 105019}
  [\href{https://arxiv.org/abs/1707.05375}{{\ttfamily 1707.05375}}].

\bibitem{Longo:2020amm}
R.~Longo and G.~Morsella, \emph{{The Massless Modular Hamiltonian}},
  \href{https://doi.org/10.1007/s00220-022-04617-1}{\emph{Commun. Math. Phys.}
  {\bfseries 400} (2023) 1181}
  [\href{https://arxiv.org/abs/2012.00565}{{\ttfamily 2012.00565}}].

\bibitem{Peschel_2003}
I.~Peschel, \emph{Calculation of reduced density matrices from correlation
  functions}, \href{https://doi.org/10.1088/0305-4470/36/14/101}{\emph{Journal
  of Physics A: Mathematical and General} {\bfseries 36} (2003) L205}.

\bibitem{Sorkin:2012sn}
R.D.~Sorkin, \emph{{Expressing entropy globally in terms of (4D)
  field-correlations}},
  \href{https://doi.org/10.1088/1742-6596/484/1/012004}{\emph{J. Phys. Conf.
  Ser.} {\bfseries 484} (2014) 012004}
  [\href{https://arxiv.org/abs/1205.2953}{{\ttfamily 1205.2953}}].

\bibitem{Saravani:2013nwa}
M.~Saravani, R.D.~Sorkin and Y.K.~Yazdi, \emph{{Spacetime entanglement entropy
  in 1 + 1 dimensions}},
  \href{https://doi.org/10.1088/0264-9381/31/21/214006}{\emph{Class. Quant.
  Grav.} {\bfseries 31} (2014) 214006}
  [\href{https://arxiv.org/abs/1311.7146}{{\ttfamily 1311.7146}}].

\bibitem{Sorkin:2016pbz}
R.D.~Sorkin and Y.K.~Yazdi, \emph{{Entanglement Entropy in Causal Set Theory}},
  \href{https://doi.org/10.1088/1361-6382/aab06f}{\emph{Class. Quant. Grav.}
  {\bfseries 35} (2018) 074004}
  [\href{https://arxiv.org/abs/1611.10281}{{\ttfamily 1611.10281}}].

\bibitem{DiGiulio:2019cxv}
G.~Di~Giulio and E.~Tonni, \emph{{On entanglement hamiltonians of an interval
  in massless harmonic chains}},
  \href{https://doi.org/10.1088/1742-5468/ab7129}{\emph{J. Stat. Mech.}
  {\bfseries 2003} (2020) 033102}
  [\href{https://arxiv.org/abs/1911.07188}{{\ttfamily 1911.07188}}].

\bibitem{Eisler:2020lyn}
V.~Eisler, G.~Di~Giulio, E.~Tonni and I.~Peschel, \emph{{Entanglement
  Hamiltonians for non-critical quantum chains}},
  \href{https://doi.org/10.1088/1742-5468/abb4da}{\emph{J. Stat. Mech.}
  {\bfseries 2010} (2020) 103102}
  [\href{https://arxiv.org/abs/2007.01804}{{\ttfamily 2007.01804}}].

\bibitem{Arias:2018tmw}
R.E.~Arias, H.~Casini, M.~Huerta and D.~Pontello, \emph{{Entropy and modular
  Hamiltonian for a free chiral scalar in two intervals}},
  \href{https://doi.org/10.1103/PhysRevD.98.125008}{\emph{Phys. Rev. D}
  {\bfseries 98} (2018) 125008}
  [\href{https://arxiv.org/abs/1809.00026}{{\ttfamily 1809.00026}}].

\bibitem{Estienne:2023ekf}
B.~Estienne, Y.~Ikhlef, A.~Rotaru and E.~Tonni, \emph{{Entanglement entropies
  of an interval for the massless scalar field in the presence of a boundary}},
   \href{https://arxiv.org/abs/2308.00614}{{\ttfamily 2308.00614}}.

\bibitem{Unruh:1976db}
W.G.~Unruh, \emph{{Notes on black hole evaporation}},
  \href{https://doi.org/10.1103/PhysRevD.14.870}{\emph{Phys. Rev. D} {\bfseries
  14} (1976) 870}.

\bibitem{Kabat:1994vj}
D.N.~Kabat and M.J.~Strassler, \emph{{A Comment on entropy and area}},
  \href{https://doi.org/10.1016/0370-2693(94)90515-0}{\emph{Phys. Lett. B}
  {\bfseries 329} (1994) 46}
  [\href{https://arxiv.org/abs/hep-th/9401125}{{\ttfamily hep-th/9401125}}].

\bibitem{Dowker:1994fi}
J.S.~Dowker, \emph{{Remarks on geometric entropy}},
  \href{https://doi.org/10.1088/0264-9381/11/4/001}{\emph{Class. Quant. Grav.}
  {\bfseries 11} (1994) L55}
  [\href{https://arxiv.org/abs/hep-th/9401159}{{\ttfamily hep-th/9401159}}].

\bibitem{Giavoni:2023cwg}
C.~Giavoni, S.~Hofmann and M.~Koegler, \emph{{Emerging entanglement on network
  histories}}, \href{https://doi.org/10.1103/PhysRevD.109.085016}{\emph{Phys.
  Rev. D} {\bfseries 109} (2024) 085016}
  [\href{https://arxiv.org/abs/2312.17313}{{\ttfamily 2312.17313}}].

\bibitem{Huerta:2022tpq}
M.~Huerta and G.~van~der Velde, \emph{{Modular Hamiltonian of the scalar in the
  semi infinite line: dimensional reduction for spherically symmetric
  regions}}, \href{https://doi.org/10.1007/JHEP06(2023)097}{\emph{JHEP}
  {\bfseries 06} (2023) 097}
  [\href{https://arxiv.org/abs/2301.00294}{{\ttfamily 2301.00294}}].

\bibitem{Boutivas:2024sat}
K.~Boutivas, D.~Katsinis, G.~Pastras and N.~Tetradis, \emph{{Entanglement
  Entropy as a Probe Beyond the Horizon}},
  \href{https://arxiv.org/abs/2407.07824}{{\ttfamily 2407.07824}}.

\bibitem{Calabrese:2012ew}
P.~Calabrese, J.~Cardy and E.~Tonni, \emph{{Entanglement negativity in quantum
  field theory}},
  \href{https://doi.org/10.1103/PhysRevLett.109.130502}{\emph{Phys. Rev. Lett.}
  {\bfseries 109} (2012) 130502}
  [\href{https://arxiv.org/abs/1206.3092}{{\ttfamily 1206.3092}}].

\bibitem{Yao_2010}
H.~Yao and X.-L.~Qi, \emph{Entanglement entropy and entanglement spectrum of
  the kitaev model},
  \href{https://doi.org/10.1103/physrevlett.105.080501}{\emph{Physical Review
  Letters} {\bfseries 105} (2010) }.

\bibitem{Pirmoradian:2023uvt}
R.~Pirmoradian and M.R.~Tanhayi, \emph{{Symmetry-Resolved Entanglement Entropy
  for Local and Non-local QFTs}},
  \href{https://arxiv.org/abs/2311.00494}{{\ttfamily 2311.00494}}.

\bibitem{Bueno:2020fle}
P.~Bueno and H.~Casini, \emph{{Reflected entropy for free scalars}},
  \href{https://doi.org/10.1007/JHEP11(2020)148}{\emph{JHEP} {\bfseries 11}
  (2020) 148} [\href{https://arxiv.org/abs/2008.11373}{{\ttfamily
  2008.11373}}].

\bibitem{Buividovich:2008gq}
P.V.~Buividovich and M.I.~Polikarpov, \emph{{Entanglement entropy in gauge
  theories and the holographic principle for electric strings}},
  \href{https://doi.org/10.1016/j.physletb.2008.10.032}{\emph{Phys. Lett. B}
  {\bfseries 670} (2008) 141}
  [\href{https://arxiv.org/abs/0806.3376}{{\ttfamily 0806.3376}}].

\bibitem{Donnelly:2011hn}
W.~Donnelly, \emph{{Decomposition of entanglement entropy in lattice gauge
  theory}}, \href{https://doi.org/10.1103/PhysRevD.85.085004}{\emph{Phys. Rev.
  D} {\bfseries 85} (2012) 085004}
  [\href{https://arxiv.org/abs/1109.0036}{{\ttfamily 1109.0036}}].

\bibitem{Casini:2013rba}
H.~Casini, M.~Huerta and J.A.~Rosabal, \emph{{Remarks on entanglement entropy
  for gauge fields}},
  \href{https://doi.org/10.1103/PhysRevD.89.085012}{\emph{Phys. Rev. D}
  {\bfseries 89} (2014) 085012}
  [\href{https://arxiv.org/abs/1312.1183}{{\ttfamily 1312.1183}}].

\bibitem{Ghosh:2015iwa}
S.~Ghosh, R.M.~Soni and S.P.~Trivedi, \emph{{On The Entanglement Entropy For
  Gauge Theories}}, \href{https://doi.org/10.1007/JHEP09(2015)069}{\emph{JHEP}
  {\bfseries 09} (2015) 069}
  [\href{https://arxiv.org/abs/1501.02593}{{\ttfamily 1501.02593}}].

\bibitem{Witten:2018zxz}
E.~Witten, \emph{{APS Medal for Exceptional Achievement in Research: Invited
  article on entanglement properties of quantum field theory}},
  \href{https://doi.org/10.1103/RevModPhys.90.045003}{\emph{Rev. Mod. Phys.}
  {\bfseries 90} (2018) 045003}
  [\href{https://arxiv.org/abs/1803.04993}{{\ttfamily 1803.04993}}].

\bibitem{Witten:2018zva}
E.~Witten, \emph{{A Mini-Introduction To Information Theory}},
  \href{https://doi.org/10.1007/s40766-020-00004-5}{\emph{Riv. Nuovo Cim.}
  {\bfseries 43} (2020) 187}
  [\href{https://arxiv.org/abs/1805.11965}{{\ttfamily 1805.11965}}].

\bibitem{Witten:2021jzq}
E.~Witten, \emph{{Why Does Quantum Field Theory In Curved Spacetime Make Sense?
  And What Happens To The Algebra of Observables In The Thermodynamic Limit?}},
   \href{https://arxiv.org/abs/2112.11614}{{\ttfamily 2112.11614}}.

\bibitem{Longo:2022lod}
R.~Longo and E.~Witten, \emph{{A note on continuous entropy}},
  \href{https://arxiv.org/abs/2202.03357}{{\ttfamily 2202.03357}}.

\bibitem{Chandrasekaran:2022eqq}
V.~Chandrasekaran, G.~Penington and E.~Witten, \emph{{Large N algebras and
  generalized entropy}},  \href{https://arxiv.org/abs/2209.10454}{{\ttfamily
  2209.10454}}.

\bibitem{Witten:2022xxp}
E.~Witten, \emph{{A Note On The Canonical Formalism for Gravity}},
  \href{https://arxiv.org/abs/2212.08270}{{\ttfamily 2212.08270}}.

\bibitem{Witten:2023qsv}
E.~Witten, \emph{{Algebras, Regions, and Observers}},
  \href{https://arxiv.org/abs/2303.02837}{{\ttfamily 2303.02837}}.

\bibitem{Kudler-Flam:2023qfl}
J.~Kudler-Flam, S.~Leutheusser and G.~Satishchandran, \emph{{Generalized Black
  Hole Entropy is von Neumann Entropy}},
  \href{https://arxiv.org/abs/2309.15897}{{\ttfamily 2309.15897}}.

\bibitem{Jafferis:2015del}
D.L.~Jafferis, A.~Lewkowycz, J.~Maldacena and S.J.~Suh, \emph{{Relative entropy
  equals bulk relative entropy}},
  \href{https://doi.org/10.1007/JHEP06(2016)004}{\emph{JHEP} {\bfseries 06}
  (2016) 004} [\href{https://arxiv.org/abs/1512.06431}{{\ttfamily
  1512.06431}}].

\bibitem{Dong:2016eik}
X.~Dong, D.~Harlow and A.C.~Wall, \emph{{Reconstruction of Bulk Operators
  within the Entanglement Wedge in Gauge-Gravity Duality}},
  \href{https://doi.org/10.1103/PhysRevLett.117.021601}{\emph{Phys. Rev. Lett.}
  {\bfseries 117} (2016) 021601}
  [\href{https://arxiv.org/abs/1601.05416}{{\ttfamily 1601.05416}}].

\bibitem{Cotler:2017erl}
J.~Cotler, P.~Hayden, G.~Penington, G.~Salton, B.~Swingle and M.~Walter,
  \emph{{Entanglement Wedge Reconstruction via Universal Recovery Channels}},
  \href{https://doi.org/10.1103/PhysRevX.9.031011}{\emph{Phys. Rev. X}
  {\bfseries 9} (2019) 031011}
  [\href{https://arxiv.org/abs/1704.05839}{{\ttfamily 1704.05839}}].

\bibitem{Leutheusser:2021qhd}
S.~Leutheusser and H.~Liu, \emph{{Causal connectability between quantum systems
  and the black hole interior in holographic duality}},
  \href{https://doi.org/10.1103/PhysRevD.108.086019}{\emph{Phys. Rev. D}
  {\bfseries 108} (2023) 086019}
  [\href{https://arxiv.org/abs/2110.05497}{{\ttfamily 2110.05497}}].

\bibitem{Leutheusser:2021frk}
S.A.W.~Leutheusser, \emph{{Emergent Times in Holographic Duality}},
  \href{https://doi.org/10.1103/PhysRevD.108.086020}{\emph{Phys. Rev. D}
  {\bfseries 108} (2023) 086020}
  [\href{https://arxiv.org/abs/2112.12156}{{\ttfamily 2112.12156}}].

\end{thebibliography}\endgroup

\end{document}